\documentclass[aps,prb,10pt,twocolumn,a4paper,superscriptaddress,nofootinbib,longbibliography]{revtex4-2}

\usepackage[colorlinks = true,
            linkcolor = blue,
            urlcolor  = blue,
            citecolor = blue,
            anchorcolor = blue,
            unicode]{hyperref}

\usepackage{graphicx}
\usepackage{amsmath}
\usepackage{amssymb}
\usepackage{amsfonts}
\usepackage{color}

\DeclareMathOperator{\sgn}{sgn}

\DeclareMathOperator{\arccosh}{arccosh}

\DeclareMathOperator{\diver}{div}

\renewcommand{\Im}{\mathop\mathrm{Im}\nolimits}

\begin{document}

\title{Enhancement of superconductivity by polarization of magnetic impurities in disordered films}

\author{Gleb~S.\ Seleznev}
\affiliation{L.~D.\ Landau Institute for Theoretical Physics RAS, 142432 Chernogolovka, Russia}
\affiliation{Moscow Institute of Physics and Technology, 141700 Dolgoprudny, Russia}

\author{Yakov~V.\ Fominov}
\affiliation{L.~D.\ Landau Institute for Theoretical Physics RAS, 142432 Chernogolovka, Russia}
\affiliation{Moscow Institute of Physics and Technology, 141700 Dolgoprudny, Russia}
\affiliation{Laboratory for Condensed Matter Physics, HSE University, 101000 Moscow, Russia}

\begin{abstract}
Dirty superconducting films with magnetic impurities can exhibit nontrivial behavior in a magnetic field that polarizes the impurity spins. As predicted by Kharitonov and Feigelman (KF) [JETP Lett.\ \textbf{82}, 421 (2005)], this polarization reduces the exchange scattering rate. Consequently, a parallel magnetic field can enhance the critical temperature $T_c$ when magnetic-field pair breaking is weak, as realized for strong spin-orbit scattering and small film thickness. Recently, Llanos \textit{et al.}\ [Nat.\ Phys.\ \textbf{22}, 856 (2026)] observed a pronounced enhancement of $T_c$ consistent with the KF theory. The same experiment also reported an enhancement of the perpendicular upper critical field $H_{c2}^{\perp}$ and a suppression of the London penetration depth $\lambda_L$ by a parallel magnetic field. These quantities were not considered in the original KF theory. To address this gap, we develop a theoretical framework based on Gor'kov's diagrammatic technique for dirty superconductors. We extend the KF theory in two experimentally relevant directions: (i)~to arbitrary temperatures $T<T_c$ and several superconducting observables, and (ii)~to arbitrary magnetic-field orientations. As a result, we demonstrate theoretically the suppression of $\lambda_L$ and the enhancement of $H_{c2}^{\perp}$ by a parallel magnetic field, in agreement with experiment.

\end{abstract}

\date{23 August 2026}

\maketitle

\section{Introduction}
\label{sec:intro}
The study of superconducting alloys containing different types of impurities has a long history and is well described in the literature \cite{deGennesBook, AbrikosovBookEng, TinkhamBook}. It was shown long ago that potential impurities do not affect the thermodynamic properties of superconductors and, in particular, do not change the critical temperature $T_c$ relative to the clean-limit value $T_{c0}$. This statement is the essence of Anderson's theorem, which relies on the preservation of time-reversal symmetry in the presence of potential impurities \cite{Abrikosov1958a,Abrikosov1959a,Anderson1959}. The situation is very different for superconducting alloys with magnetic impurities, as shown by Abrikosov and Gor'kov (AG) \cite{Abrikosov1960}. Magnetic impurities break time-reversal symmetry and therefore violate Anderson's theorem, leading to suppression of superconductivity. As their concentration increases, superconducting quantities, such as the critical temperature $T_c$, order parameter $\Delta$, and superfluid density $n_{sc}$, decrease rapidly. Above a critical concentration of magnetic impurities, superconductivity disappears completely.

A similar situation occurs when a superconductor is subjected to an external magnetic field \cite{deGennesBook, TinkhamBook, AbrikosovBookEng}. A magnetic field also breaks time-reversal symmetry and usually suppresses superconductivity. Two mechanisms are responsible for this suppression: the paramagnetic effect (PE), which originates from Zeeman splitting of the spins forming a Cooper pair, and the orbital effect (OE), caused by the increase in electron kinetic energy due to Meissner screening currents. When superconductivity is destroyed by the PE, the corresponding field is the Pauli paramagnetic limiting field $H_p$. When it is destroyed by the OE, the corresponding field is the orbital critical field $H_c$. 

Naively, one may therefore expect superconductors that contain magnetic impurities and are placed in an external magnetic field to exhibit even stronger suppression of superconductivity because these effects add up. However, Kharitonov and Feigelman (KF) pointed out an additional effect related to the polarization of the impurity spins \cite{kharitonovFeigelman}. This polarization reduces the effective exchange scattering rate on magnetic impurities and can therefore enhance superconductivity. To observe this field-induced enhancement, the PE and OE, which suppress superconductivity, must be sufficiently weak. KF therefore assumed strong spin-orbit scatterers, referred to below as spin-orbit impurities, in the system. As shown in Refs.\ \cite{Gorkov1964, Tsuneto1964, Werthamer1966, Maki1966}, spin-orbit impurities strongly reduce the PE and help overcome the Pauli limit. At the same time, to weaken the OE, they considered thin superconducting films in a field parallel to the film surface \cite{Maki1964, Larkin1965}. The resulting minimal model is a thin superconducting film containing potential, spin-orbit, and magnetic impurities in a parallel magnetic field. KF studied the dependence $T_c(H_{\parallel})$ and found that it can be enhanced. At larger magnetic fields, however, the PE and OE become important and lead to nonmonotonic behavior of $T_c(H_{\parallel})$ with a maximum at a finite field. 

Subsequent experiments searched for magnetic-field-induced enhancement of the critical temperature in thin superconducting films with magnetic impurities \cite{Garnder2011, Niwata2017}. However, the observed effect either lay outside the scope of the KF theory \cite{Garnder2011} or was weak \cite{Niwata2017}. A related enhancement effect was theoretically predicted \cite{Wei2006} and experimentally observed \cite{Rogachev2006} in a different geometry, namely, ultrathin nanowires with magnetic impurities on their surface, where a nonmonotonic magnetic-field dependence of the critical current was studied.

Recently, however, Ref.\ \cite{Llanos2026} reported a pronounced manifestation of the polarization effect in thin superconducting films. In particular, the experiment showed nonmonotonic behavior of the critical temperature in a parallel magnetic field. Importantly, this nonmonotonic behavior was found \emph{only} in magnetically doped samples, which indicates the crucial role of the polarization effect and rules out alternative mechanisms that can exist even without magnetic impurities \cite{Kogan1986.PhysRevB.34.3499, Agterberg2003, Samokhin2004, Kogan2023.PhysRevB.107.L020501}. The experimental results of Ref.\ \cite{Llanos2026} are in excellent agreement with the KF theory. 

In addition, the authors of Ref.\ \cite{Llanos2026} performed measurements at low temperatures and in a perpendicular magnetic field. They observed nonmonotonic behavior of the London penetration depth $\lambda_L$ and the perpendicular upper critical field $H^{\perp}_{c2}$. These results are beyond the original KF theory, which treats only the critical temperature. They require a theory of the enhancement effect in the developed superconducting phase, $T < T_{c}$, and for an arbitrary field orientation, $\mathbf{H} = \mathbf{H}_{\parallel} + \mathbf{H}_{\perp}$. We develop such a framework by extending the KF theory to arbitrary temperatures and field orientations. The resulting theory agrees well with the experimental results of Ref.\ \cite{Llanos2026}. 

The paper is organized as follows. In Sec.~\ref{sec:model}, we formulate the model used to study superconductivity enhancement and recall the main results of the KF theory. In Sec.~\ref{sec: Diagrammatic technique}, we derive Dyson's equation at arbitrary temperature, taking into account the polarization of impurity spins by a parallel magnetic field. We also present a general scheme for calculating superconducting observables from this equation. In Sec.~\ref{sec: superconductivity enhancement arbitrary T}, we analyze superconductivity enhancement in limiting cases. In Sec.~\ref{sec: enhancement of Hc2}, we generalize the KF theory to an arbitrary magnetic-field orientation and propose a mechanism for the enhancement of the perpendicular upper critical field by the parallel field component. In Sec.~\ref{sec: numerics}, we present numerical results showing nonmonotonic dependences of superconducting observables. In Sec.~\ref{sec: Comparison with the experimental results}, we compare our extended theory with the experiment of Ref.\ \cite{Llanos2026}. In Sec.~\ref{sec:discussion}, we discuss several aspects of the experiment and suggest directions for future studies. In Sec.~\ref{sec:conclusions}, we summarize our conclusions. Additional details are given in the Appendixes.

\section{Model}
\label{sec:model}

In this section, we present the theoretical model used to study superconductivity enhancement. We largely follow the original KF theory \cite{kharitonovFeigelman} and briefly summarize its main ideas and results.

\subsection{Hamiltonian of a disordered superconducting film in a magnetic field}
\label{sec:superconductor with magnetic impurities}

\begin{figure}[t]
 \includegraphics[width=\columnwidth]{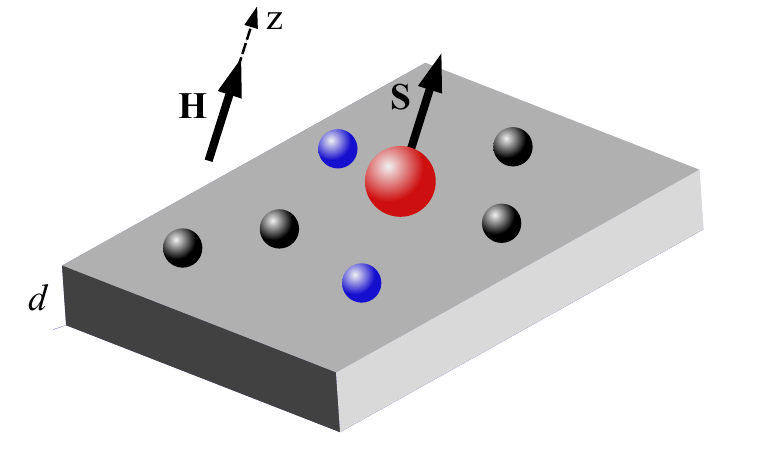}
 \caption{Disordered superconducting film of thickness $d$ containing potential (black balls), spin-orbit (blue balls), and magnetic (red balls) impurities with spin $\mathbf{S}$. The corresponding scattering rates are $\nu_{0}$, $\nu_{so}$, and $\nu_{s}$. The film is subjected to an external magnetic field $\mathbf{H}$, which polarizes the magnetic impurities. We use energy units, $\mathbf{h} = \mu_B \mathbf{H}$, where $\mu_B$ is the Bohr magneton. The field can have an arbitrary orientation, and the quantization axis $z$ is aligned with it.}
 \label{fig: Scheme}
\end{figure}

As in Ref.\ \cite{kharitonovFeigelman}, we consider a disordered superconducting film containing potential, spin-orbit, and magnetic impurities. We assume that the film is subjected to an external magnetic field $\mathbf{H}$, which may have an arbitrary orientation with respect to the film surface; see Fig.~\ref{fig: Scheme}.  

The Hamiltonian of the system can be written as
\begin{equation}\label{eq:HamiltonianFull}
H=H_{\mathrm{BCS}}+ H_{eU}+H_{eSO} + H_{eS} + H_S,   
\end{equation}
where the first term is the standard BCS Hamiltonian 
\begin{multline}
 H_{\mathrm{BCS}} =\int \Bigl\{ \psi^{\dag}_{\alpha} \left[ \frac{(\mathbf{p}-e\mathbf{A}/c )^2}{2m}-E_F\right] \psi_{\alpha} \\ +\frac{\lambda}{2} \psi^{\dag}_{\alpha} \psi^{\dag}_{\beta} \psi_{\beta} \psi_{\alpha}
   - \psi^{\dag}_{\alpha} h(\sigma_{z})_{\alpha \beta} \psi_{\beta} \Bigl\} d^3 \mathbf{r},
\end{multline}
which includes the terms corresponding to the orbital (proportional to the vector potential $\mathbf{A}$) and paramagnetic (proportional to the Zeeman field $h=\mu_B H$) effects of the external magnetic field. Here, $e$ is the electron charge, $c$ is the speed of light, $\mu_B$ is the Bohr magneton, and we put $\hbar=1$. The quantities $E_F$ and $m$ denote the Fermi energy and the electron effective mass, respectively. Thus, the normal-state electronic structure is modeled as a single isotropic three-dimensional band with a spherical Fermi surface. The indices $\alpha$ and $\beta$ label electron spin, and $\hat{\sigma}_z$ is the third Pauli matrix in spin space. Note that BCS pairing involves electrons with opposite spins, $\beta \neq \alpha$.

The second, third, and fourth terms in Eq.\ \eqref{eq:HamiltonianFull} describe the interaction of electrons with different types of impurities. The interaction with potential impurities is described by the second term:
\begin{equation}\label{eq: HeU}
 H_{eU} = \int \Bigl\{  \sum_{a} \psi^{\dag}_{\alpha} u_{0} \delta_{\alpha \beta} 
  \delta(\mathbf{r}-\mathbf{R}_{a})  \psi_{\beta}  \Bigl\} d^3 \mathbf{r},
\end{equation}
where $u_0$ is the scattering amplitude of the potential impurities.

The third term in Eq.\ \eqref{eq:HamiltonianFull} describes the interaction of electrons with spin-orbit impurities:
\begin{equation}\label{eq: HeSO}
 H_{eSO} =  \int \Bigl \{\sum_{b}\psi^{\dag}_{\alpha}(\mathbf{r})  v_{\alpha \beta}(\mathbf{r}-\mathbf{R}_b, \mathbf{r}'-\mathbf{R}_b)
   \psi_{\beta}(\mathbf{r}')  \Bigl\} d^3 \mathbf{r} d^3 \mathbf{r}',
\end{equation}
where $v_{\alpha \beta}(\mathbf{r},\mathbf{r}')$ is the spin-orbit impurity amplitude in coordinate representation. In momentum space, it has the form
$v_{\alpha \beta}(\mathbf{p},\mathbf{p}')=i \left(v_{so}/p_F^2 \right) ([\mathbf{p} \times\mathbf{p}'] \cdot \boldsymbol\sigma_{\alpha \beta})$.

The interaction of electrons with magnetic impurities is given by the fourth term in Eq.\ \eqref{eq:HamiltonianFull}: 
\begin{equation}\label{eq: HeS}
 H_{eS} = \int \Bigl\{  \sum_{c} \psi^{\dag}_{\alpha} J (\mathbf{S}_{c} \cdot \boldsymbol\sigma_{\alpha\beta})
  \delta(\mathbf{r}-\mathbf{R}_{c})  \psi_{\beta}  \Bigl\} d^3 \mathbf{r},
\end{equation}
where $J$ is the exchange coupling constant and $\mathbf{S}_c$ is the spin of the magnetic impurity.

Finally, the last term in Eq.\ \eqref{eq:HamiltonianFull} describes the internal spin dynamics and gives the polarization of magnetic-impurity spins by the applied magnetic field: 
\begin{equation}\label{eq: Hs}
 H_{S}= - \sum_c \omega_s S^{z}_{c},
\end{equation}
where $\omega_s=g_s h$ is the Zeeman energy and $g_s$ is the $g$ factor of the magnetic impurities.

\subsection{Assumptions}\label{sec: asummptions}

We now discuss the assumptions under which we analyze the Hamiltonian in Eq.\ \eqref{eq:HamiltonianFull}.

\subsubsection{Disorder}

First, we state the assumptions that allow us to use the standard Gor'kov diagrammatic technique for dirty superconductors.
 
 We assume that the different impurity types are mutually independent and uniformly distributed throughout the sample, with concentrations $n_0$ (potential), $n_{so}$ (spin-orbit), and $n_s$ (magnetic). This assumption differs from the KF theory, where some impurity types were not independent; see Appendix~\ref{appensix: interferential terms}. However, interference between different types of disorder did not affect the polarization effect in the KF theory \cite{kharitonovFeigelman}. We therefore use the simpler model of independent impurities. 

 Regarding impurity strength, we work in the Born approximation. In particular, this requires $\zeta_s=1/2\pi N_F |J|(S+1) \gg 1$, where $\zeta_s$ is the inverse Born parameter for exchange scattering and $N_F$ is the density of states at the Fermi energy. We also neglect the Kondo effect by considering temperatures $T>T_K \propto E_F e^{-1/N_F|J|}$, where $T_K$ is the Kondo temperature and is exponentially small in the Born approximation. In the low-temperature limit denoted below by $T=0$, we still assume $T>T_K$.
 
Finally, we assume $p_F l \gg 1$, where $p_{F}$ is the Fermi momentum and $l$ is the mean free path, and focus on the ``dirty'' limit with respect to potential and spin-orbit scattering. This implies the hierarchy
 \begin{equation}\label{eq: dirty condition}
\nu_s, T_{c0}, h\ll \nu_{so} \ll \nu_0,
 \end{equation}
where $\nu_{0} = 2 \pi N_{F} n_0 u_0^2$, $\nu_{so} = (4\pi/3) N_{F} n_{so} v_{so}^2$, and $\nu_{s} = 2 \pi N_{F} n_s J^2 S (S+1)$ are the scattering rates for potential, spin-orbit, and magnetic impurities, respectively. In our analysis, $\nu_s$, $T_{c0}$, and $h$ may be comparable.

\subsubsection{Conditions for observing the enhancement effect}

To observe superconductivity enhancement, the PE and OE must be weak compared with the polarization effect. This requires comparing the corresponding pair-breaking parameters [see Eq.\ \eqref{eq: pair-breaking parameters} below] with the energy gain from polarization, which is of order $\nu_s$. The PE is weak when spin-orbit scattering is strong:
\begin{equation}\label{eq: weakness of PE}
 h'^2/\nu_{so} \ll \nu_{s},
\end{equation}
where $h'$ is the screened Zeeman field, which includes the field produced by the polarized impurity spins: 
\begin{equation}\label{eq: h'}
h' = h - \nu_s \zeta_s \langle S_z \rangle/S,
\end{equation}
where $\langle S_z \rangle$ is the thermodynamic average of the impurity-spin projection on the $z$ axis [given by Eq.\ \eqref{eq: Sz} below]. 

To make the OE weak, we assume that the film thickness $d$ is small:
\begin{equation}\label{eq: weakness of OE}
(p_{F}d)^2h^2/\nu_0  \ll \nu_s,  
\end{equation}
while $d/l \gg 1$. Moreover, we assume $d\ll \lambda_L$ and $d\ll l_{H_{\parallel}}=\sqrt{c/eH_{\parallel}}$, where $l_{H_{\parallel}}$ is the magnetic length corresponding to the parallel magnetic field. The first condition means that the parallel field is not screened by supercurrents, whereas the second means that vortices do not fit within the film thickness.

We emphasize that in the experimentally relevant case $h \sim \nu_s \sim T_{c0}$, our Eqs.\ \eqref{eq: weakness of PE}
and \eqref{eq: weakness of OE} reproduce the corresponding conditions of the KF theory \cite{kharitonovFeigelman}:
\begin{equation}
1 \ll \zeta_s^2 \ll \nu_{so}/\nu_{s} , \quad  (p_{F}d)^2\ll \nu_0/T_{c0}. 
\end{equation}

Under the assumptions in Eqs.\ \eqref{eq: weakness of PE} and \eqref{eq: weakness of OE}, magnetic-field-induced superconductivity enhancement can be observed. Below, we use these conditions when we neglect the PE and OE and focus only on the polarization effect. We include the PE and OE when studying the competition between polarization and magnetic-field pair breaking. 

\subsection{Abrikosov--Gor'kov and Kharitonov--Feigelman theories}

Before proceeding to the general analysis, we briefly recall the main result of the KF theory. We first summarize the well-known AG theory \cite{Abrikosov1960}, which describes a superconducting film with magnetic impurities in the absence of an external magnetic field, $h=0$.

In the AG theory, the absence of an external magnetic field means that the impurity spins are unpolarized. The only effect of magnetic impurities is therefore the suppression of superconductivity by exchange scattering. In particular, the critical temperature is determined by
 \begin{equation}\label{eq: Tc AG theory}
   \ln{\frac{T_{c0}}{T}} = \pi T \sum_{\varepsilon_{n}} \left(\frac{1}{|\varepsilon_{n}|} - \frac{1}{|\varepsilon_{n}|+\nu_s}\right),
\end{equation}
where $\varepsilon_{n} = 2 \pi T (n +1/2)$ is a fermionic Matsubara frequency. The solution of Eq.\ \eqref{eq: Tc AG theory} defines the dependence $T_c=T_{\mathrm{AG}}(\nu_s)$, which decreases monotonically with increasing $\nu_s$ (or, equivalently, $n_s$). More generally, $\nu_s$ suppresses other superconducting characteristics at any temperature as well.

In the KF theory \cite{kharitonovFeigelman}, the authors considered an additional effect caused by a magnetic field parallel to the film surface, $h_{\parallel}\neq0$. As discussed above, this field polarizes the impurity spins and reduces the effective exchange scattering rate. The critical temperature is then determined by
\begin{equation}\label{eq: Tc KF theory}
   \ln{\frac{T_{c0}}{T}} = \pi T \sum_{\varepsilon_{n}} \left(\frac{1}{|\varepsilon_{n}|} - C_0 (\varepsilon_{n})\right),
\end{equation}
where $C_0 (\varepsilon_{n})$ is a Cooperon (strictly speaking, this is the singlet Cooperon component) \cite{Akkermans2007, Rammer1986.RevModPhys.58.323, LevitovShytovBook}  which satisfies the following equation \cite{kharitonovFeigelman}:
\begin{equation}\label{eq: Cooperon}
    (|\varepsilon_{n}| +\hat{\nu}_{ex} + \gamma^{\parallel}_{PE}  + \gamma^{\parallel}_{OE} )C_0(\varepsilon_{n})=1,
\end{equation}
where $\gamma^{\parallel}_{PE}$ and $\gamma^{\parallel}_{OE}$ are the pair-breaking parameters associated with the PE and OE of the parallel magnetic field, respectively. The quantity $\hat{\nu}_{ex}$ is the effective exchange scattering rate due to magnetic impurities. In general, $\hat{\nu}_{ex}$ is a nonlocal operator in Matsubara frequency because the impurity spins are dynamical in a magnetic field; the hat denotes this operator character. It becomes a simple number in two limiting cases: (i)~the unpolarized limit, $h_{\parallel}=0$, and (ii)~the fully polarized limit, symbolically denoted by $h_{\parallel}\rightarrow\infty$, in which all impurity spins are aligned with the field:
\begin{equation}\label{eq: nu ex}
    \nu_{ex} = \begin{cases}
        \nu_s,  &h_{\parallel} =0, \\
        \nu_{\infty} = \nu_s S/(S+1) ,  &h_{\parallel} \rightarrow \infty.
    \end{cases}
\end{equation}
We emphasize that the unpolarized limit is applicable when $h \ll T_{c0}$, whereas the fully polarized limit implies that $h \gg T_{c0}$. Moreover, the fully polarized limit does not necessarily imply strong PE or OE. Under the conditions in Eqs.\ \eqref{eq: weakness of PE} and \eqref{eq: weakness of OE}, both effects remain weak. Thus, the PE and OE can be neglected in the fully polarized limit when the film is thin and spin-orbit scattering is strong, respectively. 

Therefore, KF predicted \cite{kharitonovFeigelman} that if the PE and OE are sufficiently weak, the polarization effect dominates and superconductivity is enhanced relative to the case $h_{\parallel}=0$. Formally, this requires the $h_{\parallel}$-dependent Cooperon to satisfy
\begin{equation}\label{eq: condition for Tc enhancement}
\sum_{\varepsilon_{n}} \left(\frac{1}{|\varepsilon_{n}|+\nu_s} - C_0 (\varepsilon_{n})\right) < 0,
\end{equation} 
which corresponds to an enhancement of the critical temperature relative to the AG value [see Eqs.\ \eqref{eq: Tc AG theory} and \eqref{eq: Tc KF theory}]. However, Eq.\ \eqref{eq: condition for Tc enhancement} is inevitably violated at sufficiently large magnetic fields because the PE and OE grow without bound, whereas the exchange scattering rate can decrease only down to $\hat{\nu}_{ex} \rightarrow \nu_{\infty}$ as $h_{\parallel} \rightarrow \infty$. Thus, within the KF theory, $T_c(h_{\parallel})$ either decreases monotonically, if Eq.\ \eqref{eq: condition for Tc enhancement} is violated for all $h_{\parallel}$, or is nonmonotonic, with a maximum at a finite magnetic field.

Thus, the KF theory extends the AG theory by capturing the polarization of magnetic impurities in a finite magnetic field. However, the analysis of Ref.\ \cite{kharitonovFeigelman} describes only the manifestation of the polarization effect in the critical temperature $T_{c}$ and is restricted to a parallel magnetic field, $\mathbf{h} = \mathbf{h}_{\parallel}$. In the next sections, we show how to extend the KF theory to other superconducting observables and an arbitrary magnetic-field orientation.

\section{Diagrammatic technique}\label{sec: Diagrammatic technique}
In this section, we use a diagrammatic technique to derive Dyson's equation for the Green's function at an arbitrary temperature $T<T_c$, including the effect of impurity-spin polarization. We also describe a general scheme for calculating the magnetic-field dependence of superconducting observables. We first consider a parallel magnetic field, $\mathbf{h}=\mathbf{h}_{\parallel}$, and postpone the discussion of perpendicular-field effects until Sec.\ \ref{sec: enhancement of Hc2}.

\subsection{Dirty superconductor in a magnetic field}\label{sec: dirty superconductor in the magnetic field}

To describe a dirty superconducting film at arbitrary temperature $T<T_c$, we use Gor'kov's diagrammatic technique in the Matsubara representation \cite{AGDBookEng, LevitovShytovBook, Maki1969}. The Bogoliubov--de~Gennes (BdG) Hamiltonian of the electron subsystem [Eq.\ \eqref{eq:HamiltonianFull} without the $H_S$ term] can be written as a $4\times4$ matrix in Gor'kov--Nambu (GN) and spin spaces:
\begin{widetext}
\begin{equation}\label{eq: GN Hamiltonian}
    \check{H}^{(\mathrm{BdG})} = \begin{pmatrix}
     H_0 \hat{\sigma}_{0}+\left(\mathbf{U}_{so}+\mathbf{U}_{s} \right)\cdot \hat{\boldsymbol\sigma} - h_{\parallel} \hat{\sigma}_{z} & \Delta \hat{\sigma}_{0} \\
    \Delta^{*} \hat{\sigma}_0 & - H^*_0 \hat{\sigma}_{0} - \left(\mathbf{U}_{so}-\mathbf{U}_{s} \right) \cdot \hat{\boldsymbol\sigma} - h_{\parallel} \hat{\sigma}_{z}
    \end{pmatrix}.
\end{equation}
\end{widetext}
Here, $H_0=(\mathbf{p}-e\mathbf{A}/c)^2/2m-E_F+U_0$, whereas $U_0$, $\mathbf{U}_{so}$, and $\mathbf{U}_{s}$ are the potentials produced by potential, spin-orbit, and magnetic impurities, respectively. The matrix $\hat{\sigma}_i$ is the $i$th Pauli matrix in spin space, and the Hamiltonian is written explicitly in GN space. The Green's function $\check g$ is determined by
\begin{equation}\label{eq: Green function}
    \left(i \varepsilon_{n} \hat{\tau}_{0} \hat{\sigma}_0 - \check{H}^{(\mathrm{BdG})}\right) \check{g} = \check{1},
\end{equation}
where $\hat{\tau}_{i}$ is the $i$th Pauli matrix in the GN space. 

The Green's function $\check g$ in Eq.\ \eqref{eq: Green function} corresponds to one disorder realization. To obtain disorder-averaged quantities, one must average over disorder and introduce $\check G=\langle \check g\rangle_{\mathrm{dis}}$. As shown in Ref.\ \cite{Maki1969}, the elements of $\check G$ can be sought in the form
\begin{multline}\label{eq:Green function zero perp field}
    \hat{G}^{\mathbf{q}_{s}}_{\pm} ( \varepsilon_{n},\mathbf{p}) \\= \frac{(i\tilde{\varepsilon}_{n} \mp \tilde{h}_{\parallel} +\mathbf{v}_{\mathbf{p}} \cdot \mathbf{q}_s ) \hat{\tau}_0 + \xi_{\mathbf{p}} \hat{\tau}_z - \tilde{\Delta}_{\pm} \hat{\tau}_{+} - \tilde{\Delta}_{\pm}^{*} \hat{\tau}_{-}  }
    {(i\tilde{\varepsilon}_{n} \mp \tilde{h}_{\parallel}  + \mathbf{v}_{\mathbf{p}} \cdot \mathbf{q}_s)^2 - \xi_{\mathbf{p}}^2 - \tilde{\Delta}_{\pm}^2},
\end{multline}
where the Green's function is written as a function of the momentum $\mathbf{p}$, which corresponds to the Fermi velocity $\mathbf{v}_{\mathbf{p}}$, and the Matsubara frequency $\varepsilon_n$. The latter enters Eq.\ \eqref{eq:Green function zero perp field} through the renormalized values of the Matsubara frequency $ \tilde{\varepsilon}_n$,  spin-dependent superconducting gaps $\tilde{\Delta}_{\pm}$, and magnetic field $\tilde{h}_{\parallel}$, all of which depend on $\varepsilon_n$. In general, these quantities do not coincide with the corresponding bare Matsubara frequency $ \varepsilon_n$, superconducting gap $\Delta$, and magnetic field $h_{\parallel}$.

The Green's function $\check {G}$ has a nontrivial matrix structure. To indicate its spin structure in the presence of the Zeeman field, we use the signs $\pm$, with plus and minus corresponding to positive and negative spin projections on the $z$ axis, respectively. Thus, $\check{G}$ is diagonal in spin space [see Eq.\ \eqref{eq:Green function zero perp field}] but contains both $\hat{\sigma}_0$ and $\hat\sigma_z$ components when $h_{\parallel} \neq 0$. The renormalized spin-dependent gaps $\tilde{\Delta}_{\pm}$ depend on the spin projection, whereas the renormalized magnetic field $\tilde{h}_{\parallel}$ enters the two spin components with opposite signs. To specify the structure in GN space, we also introduce $\hat{\tau}_+ = (\hat{\tau}_x + i \hat{\tau}_y)/2$ and $\hat{\tau}_- = (\hat{\tau}_x - i \hat{\tau}_y)/2$. 

The Green's function in Eq.\ \eqref{eq:Green function zero perp field} also includes the vector potential $\mathbf{A}$ explicitly. It enters through $\xi_{\mathbf{p}} = p^2/2m + e^2 A^2/2mc^2 - E_F$ and $\mathbf{q}_s = (e/c) \mathbf{A}$, which is the field-induced electron momentum. In general, the form in Eq.\ \eqref{eq:Green function zero perp field} is not applicable because the vector potential varies in space and does not commute with the momentum operator $\hat{\mathbf{p}}$. However, when the characteristic spatial scale of the problem is smaller than the magnetic length $l_H = \sqrt{c/eH}$, one can use Eq.\ \eqref{eq:Green function zero perp field} as if the vector potential were constant, with subsequent averaging over the relevant scale \cite{Gorkov1959, Maki1964, Maki1969}. In our case, this requires $d \ll l_{H_{\parallel}}$ and averaging over the film thickness. Equation~\eqref{eq:Green function zero perp field} also assumes the gauge $\diver \mathbf{A} = 0$.

\begin{figure*}[t]
\begin{center}
\includegraphics[width=17.5cm]{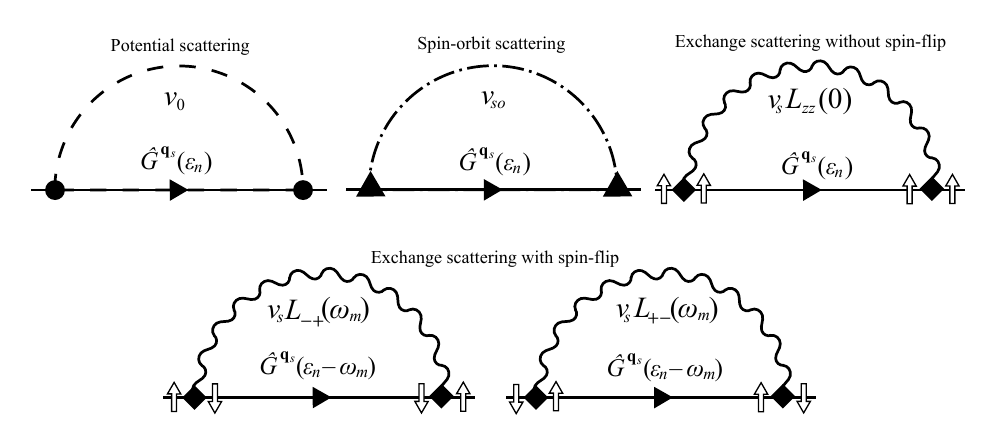}
\caption{Self-energy diagrams in the self-consistent Born approximation, calculated using the exact Green's function $\hat{G}^{\mathbf{q}_s}(\varepsilon_n, \mathbf{p})$ from Eq.\ \eqref{eq:Green function zero perp field}. The diagrams describe scattering by potential impurities (scattering rate $\nu_0$), spin-orbit impurities (scattering rate $\nu_{so}$), and magnetic impurities (scattering rate $\nu_s$). Exchange scattering by magnetic impurities is divided into two groups: processes without and with spin flips. The impurity lines in the top row, which describe potential scattering, spin-orbit scattering, and exchange scattering without spin flips (proportional to $L_{zz}$), carry zero Matsubara frequency, $\omega_m = 0$. The impurity lines in the bottom row, which describe spin-flip processes and are proportional to $L_{-+}$ and $L_{+-}$, carry nonzero Matsubara frequency, $\omega_m \neq 0$ [see Eq.\ \eqref{eq: correlators}]. This nonzero frequency transfer reflects the thermodynamics of the impurity spins in a magnetic field.}
\label{fig: Diagramms}
\end{center}
\end{figure*}

 Finally, we relate the renormalized Matsubara frequency $\tilde{\varepsilon}_n$, spin-dependent superconducting gaps $\tilde{\Delta}_{\pm}$, and renormalized magnetic field $\tilde{h}_{\parallel}$ to their bare values $ \varepsilon_n$, $\Delta$, and $h_{\parallel}$, respectively. For this purpose, we calculate the self-energy diagrams with different impurity lines and solve the corresponding Dyson's equation in the self-consistent Born approximation \cite{Abrikosov1958a, Abrikosov1960, Gorkov1964, Tsuneto1964, Maki1969}. The self-energy diagrams are shown in Fig.~\ref{fig: Diagramms}. The potential- and spin-orbit-scattering diagrams have the standard form \cite{Abrikosov1958a, Gorkov1964, Tsuneto1964}, whereas the exchange-scattering diagrams require additional discussion.

\subsection{Spin correlation functions and exchange scattering diagrams}\label{sec: spin correlation functions}
A special feature of the present system is that the impurity lines describing exchange scattering in Fig.~\ref{fig: Diagramms} depend on the transferred Matsubara frequency $\omega_m$. This dependence arises from the thermodynamics of impurity spins polarized by the magnetic field. In the absence of the Kondo effect, this spin dynamics is fully described by $H_S$ in Eq.\ \eqref{eq: Hs}. The corresponding statistical-mechanics problem gives the impurity-spin correlation functions \cite{MahanBook}
 \begin{gather}
    L_{zz}( \omega_{m}) = \int^{1/T}_{0} \langle S_{z} (\tau) S_{z}(0) \rangle e^{i \omega_{m} \tau} d\tau =  \frac{\langle S^{2}_{z} \rangle \delta_{m0}}{T}, \notag \\
    L_{-+}(\omega_{m}) = \int^{1/T}_{0} \langle S_{-} (\tau) S_{+}(0) \rangle e^{i \omega_{m} \tau} d\tau = \frac{2 \langle S_{z} \rangle} {i \omega_m + \omega_s}, \notag \\
    L_{+-}( \omega_{m}) = \int^{1/T}_{0} \langle S_{+} (\tau) S_{-}(0) \rangle e^{i \omega_{m} \tau} d\tau = \frac{2\langle S_{z} \rangle} {-i \omega_m + \omega_s}, \label{eq: correlators}
\end{gather}
where $\omega_m = 2 \pi m T$ is the bosonic Matsubara frequency transferred by the impurity line. 

In the expressions above, $\langle \dots \rangle$ denotes thermodynamic averaging over the Gibbs ensemble. This averaging gives
\begin{gather}
    \langle S_{z}\rangle = \left(S + \frac{1}{2}\right)\coth \left[{\left(S + \frac{1}{2}\right) \frac{\omega_{s}}{T}}\right] - \frac{1}{2} \coth \frac{\omega_s}{2 T}, \notag \\
    \langle S_{z}^{2} \rangle =  S (S+1) - \langle S_z \rangle \coth({\omega_{s}/{2 T}}). 
    \label{eq: Sz}
\end{gather} 

The different impurity-spin correlation functions in Eq.\ \eqref{eq: correlators} enter different self-energy diagrams for exchange scattering by magnetic impurities; see Fig.~\ref{fig: Diagramms}. We divide the exchange-scattering processes, and the corresponding diagrams, into two groups: processes with and without electron spin flips. Note that an electron spin flip is accompanied by the corresponding change of the magnetic impurity spin projection.

The diagram describing scattering by a magnetic impurity without an electron spin flip is shown in the top row of Fig.~\ref{fig: Diagramms}. The corresponding impurity line contains $L_{zz} (\omega_m) \propto \delta_{m0}$ and therefore does not transfer Matsubara frequency; see Eq.\ \eqref{eq: correlators}. As a result, the exchange scattering rate $\nu_{z}$ for processes without spin flips is only weakly sensitive to the magnetic field. In particular, it does not vanish in the fully polarized limit: $\nu_{z} \rightarrow \nu_\infty$ as $h_{\parallel} \rightarrow \infty$.

The diagrams describing exchange scattering with spin flips are shown in the bottom row of Fig.~\ref{fig: Diagramms}. Their impurity lines contain $L_{-+} (\omega_m)$ and $L_{+-} (\omega_m)$ and therefore carry nonzero frequency $ \omega_m$. This reflects the thermodynamics of impurity spins in a magnetic field. Consequently, the spin-flip exchange scattering rate $\nu_{\perp}$ is much more sensitive to the magnetic field. In particular, spin-flip processes are frozen out in the fully polarized limit: $\nu_{\perp} \rightarrow 0$ as $h_{\parallel} \rightarrow \infty$.

Finally, the effective exchange scattering rate is the sum of the exchange scattering rates without and with spin-flips:
\begin{equation}\label{eq: nu ex def}
    \hat{\nu}_{ex} = \nu_{z} + \hat{\nu}_{\perp}.
\end{equation}
As in the KF theory, these scattering rates are operators in the general case, as indicated by hats [their explicit forms are explained in the next subsection by Eqs.\ \eqref{eq:symmetric part} and \eqref{eq: spin-flip rate}]. They reduce to numbers in two limiting cases: (i)~the unpolarized limit, $h_{\parallel} = 0$, where $\nu_{z} = \nu_{s}/3$ and $\nu_{\perp} = 2\nu_{s}/3$, and (ii)~the fully polarized limit, $h_{\parallel} \rightarrow \infty$, where $\nu_{z}=\nu_{\infty}$ and $\nu_{\perp}=0$.

\subsection{Dyson's equation and general scheme of calculations}

As mentioned above, to find the Green's function, one must relate the renormalized quantities indicated by tildes in Eq.\ \eqref{eq:Green function zero perp field} to their bare values. This can be done self-consistently by solving Dyson's equation:
\begin{equation}\label{eq: Dyson}
    \bigl(\hat{G}^{\mathbf{q}_s}_{ \pm}\bigr)^{-1} = \bigl(\hat{G}^{\mathbf{q}_s}_{0, \pm}\bigr)^{-1} - \hat{\Sigma}^{\mathbf{q}_s}_ {\pm},
\end{equation}
Here, $\hat{G}_{0,\pm}^{ \mathbf{q}_s }$ is the bare Green's function. It has the same form as the exact Green's function in Eq.\ \eqref{eq:Green function zero perp field}, but with the bare values $ \varepsilon_n$, $\Delta$, and $h_{\parallel}$. The self-energy $\hat{\Sigma}^{\mathbf{q}_s}_{ \pm}$ contains the diagrams in Fig.~\ref{fig: Diagramms} and must be calculated using the exact Green's function $\hat{G}^{\mathbf{q}_s}_{\pm}$. 

Assuming strong spin-orbit scattering, we obtain the following equation from Dyson's equation (see Appendix~\ref{Appendix: Dyson's equation} for  details):
\begin{gather}\label{eq:symmetric part}
  \frac{\varepsilon_{n}}{\Delta} = u_{s} - \frac{\gamma_{\parallel} + \hat{\nu}_{ex}}{\Delta} \frac{u_{s}}{\sqrt{1 + u^2_{s}}} 
  , \\
  u_{a} = \frac{3 i h'_{\parallel}}{2\nu_{so}} \sqrt{1 + u_{s}^2}. \label{eq:asymmetric part}
\end{gather}
Equations~\eqref{eq:symmetric part} and \eqref{eq:asymmetric part} are written for the symmetric part $u_s$ and antisymmetric part $u_a$ of $u_{\pm} = (\tilde{\varepsilon}_n \pm i \tilde{h}_{\parallel})/\tilde{\Delta}_{\pm}$, so that $u_{\pm}=u_s \pm u_a$. In the strong spin-orbit-scattering limit considered above, $u_a$ is parametrically smaller than $u_s$ (by a factor $h'_{\parallel}/ \nu_{so}$) and thus will be neglected below.

In Eq.\ \eqref{eq:symmetric part} $\gamma_{\parallel}$ is the pair-breaking parameter of the parallel magnetic field, which contains two terms related to the PE and OE:
\begin{gather}\label{eq: pair-breaking parameters}
\gamma_{\parallel} = \gamma^{\parallel}_{PE}+\gamma^{\parallel}_{OE}, \\
 \gamma^{\parallel}_{PE}= 3 h'^2_{\parallel}/2\nu_{so}, \quad \gamma^{\parallel}_{OE}= 2 (p_{F}d)^2 h_{\parallel}^2/9 \nu_0. \notag
\end{gather}
We note that Eq.\ \eqref{eq: Cooperon} contains $\gamma^{\parallel}_{PE}$ and $\gamma^{\parallel}_{OE}$ given by Eq.\ \eqref{eq: pair-breaking parameters}.

Finally, Eq.\ \eqref{eq:symmetric part} contains the exchange scattering rate $\hat{\nu}_{ex}$ given by Eq.\ \eqref{eq: nu ex def} with
\begin{gather}
   \nu_z= \nu_s \langle S_z^2\rangle/S (S+1),\\
   \frac{\hat{\nu}_{\perp}}{\Delta} \frac{u_{s}}{\sqrt{1 + u^2_{s}}}= \frac{\nu_{s} \langle S_z \rangle T}{ \Delta S (S+1)}  \sum \limits_{\omega_{m}} \frac{ \omega_{s} }{\omega_s^2 + \omega_{m}^2} \frac{ u_{s} + \bar{u}_s}{\sqrt{1 + \bar{u}_{s}^2}}, \label{eq: spin-flip rate}
   \end{gather}
where $u_s = u_s(\varepsilon_n)$ and $\bar{u}_s = u_s (\varepsilon_n - \omega_m)$. The frequency shift $\varepsilon_n - \omega_m$ in the argument of $\bar u_s$ explicitly shows that $\hat{\nu}_{ex}$ is a nonlocal operator in Matsubara frequency. 

To calculate superconducting observables, we use the following general scheme \cite{Maki1969}. First, we solve Eq.\ \eqref{eq:symmetric part} to obtain the dependence $u_{s} (\varepsilon_n)$. Next, we express the observable in terms of the exact Green's function $\hat{G}^{\mathbf{q}_s}_{\pm}$, for example using the Kubo formula for linear response \cite{LevitovShytovBook, MahanBook}. Finally, we rewrite the result in terms of $u_s$ and substitute the solution obtained in the first step. This procedure gives the following expressions for the self-consistency equation, the superfluid density $n_{sc}$, and the density of states $N(E)$:

\begin{gather}
    \ln\frac{T_{c0}}{T} = \pi T \sum_{\varepsilon_{n}} \left(\frac{1}{|\varepsilon_n|} -\frac{1}{\Delta \sqrt{1+u_s^2(\varepsilon_n)}} \right), \label{eq: self-consistency}
    \\
    \frac{n_{sc}}{n}  = \frac{2\pi T}{\nu_0} \sum_{\varepsilon_{n}} \frac{1}{ 1+u_{s}^2(\varepsilon_n)}, \label{eq: nsc} \\
    N(E)= N_F \Im \biggl(\frac{ u_s ( \varepsilon_n)}{\sqrt{1 + u^2_{s}(\varepsilon_n)}}\biggl|_{\varepsilon_n \mapsto -iE} \biggl), \label{eq: spectral gap}
\end{gather}
where $n$ is the total electron density, and the substitution $\varepsilon_n \mapsto -i E$ in Eq.\ \eqref{eq: spectral gap} implies analytic continuation to the real energy $E$.

Equations~\eqref{eq: self-consistency}--\eqref{eq: spectral gap} exactly reproduce the corresponding expressions of the AG theory~\cite{Abrikosov1960}. However, Eq.~\eqref{eq:symmetric part} for $u_s$ is more complicated than its AG counterpart; compare Eq.~\eqref{eq:symmetric part} with Eq.~\eqref{eq: u at zero field} below.

Following the general scheme described
above, we solve Eq.\ \eqref{eq:symmetric part} and substitute the resulting dependence $u_s (\varepsilon_n)$ into Eqs.\ \eqref{eq: self-consistency}--\eqref{eq: spectral gap} to obtain $\Delta$, $n_{sc}$, and $N(E)$, respectively.

\subsection{Dyson's equation: correspondence to the AG and KF theories}
Before calculating superconducting observables with this scheme, we consider limiting cases of Eq.\ \eqref{eq:symmetric part} and show how it reduces to the AG and KF theories. 

First, Eq.\ \eqref{eq:symmetric part} is nonlocal in Matsubara frequency because of the $\hat{\nu}_{\perp}$ term [see Eq.\ \eqref{eq: spin-flip rate}] and is nonlinear in $u_s$. Therefore, in the most general case it can be solved only numerically. In several limiting cases, however, Eq.\ \eqref{eq:symmetric part} simplifies and yields analytical results.

One of these limiting cases is the situation when the temperature is near the critical value, $T \rightarrow T_{c}$. In this limit, the order parameter is small $\Delta \rightarrow 0 $, which implies that $u_{s} \gg 1$. Equation \eqref{eq:symmetric part} then reproduces the equation determining the Cooperon $C_0 (\varepsilon_n)$ in the KF theory [see Eq.\ \eqref{eq: Cooperon}] with $C_0 (\varepsilon_n) = |u_s^{-1} (\varepsilon_n)|/\Delta$.

Analytical results can also be obtained at arbitrary temperature in the unpolarized and fully polarized limits introduced above. In the unpolarized limit where $\omega_s  =0$, only the zeroth harmonic $\omega_{m} = 0 $ in the sum contributes, which leads to 
\begin{equation}\label{eq: u at zero field}
\frac{\varepsilon_{n}}{\Delta} = u_{s} - \frac{\nu_{s}}{\Delta} \frac{u_{s}}{\sqrt{1 + u^2_{s}}}, \quad h_{\parallel}=0,
\end{equation}
which exactly reproduces Dyson's equation obtained in the AG theory \cite{Abrikosov1960}. 

In the opposite limit of full polarization, $h_{\parallel} \rightarrow \infty$, the sum in Eq.~\eqref{eq: spin-flip rate}, which describes spin-flip processes, vanishes. Thus, Eq.~\eqref{eq:symmetric part} takes the form
\begin{equation} \label{eq: u at infinity field}
    \frac{\varepsilon_{n}}{\Delta} = u_{s}  - \frac{\gamma_{\parallel} +\nu_{\infty}}{\Delta} \frac{u_{s}}{\sqrt{1 + u^2_{s}}}, \quad h_{\parallel}\rightarrow \infty.
\end{equation}
If, in addition, the PE and OE are negligibly small, $\gamma_{\parallel}=0$, then this equation reduces to the AG-like form
\begin{equation}\label{eq: us fully polarized}
    \frac{\varepsilon_{n}}{\Delta} = u_{s}  - \frac{\nu_{\infty}}{\Delta} \frac{u_{s}}{\sqrt{1 + u^2_{s}}}, \quad h_{\parallel}\rightarrow \infty,
\end{equation}
with the effective exchange scattering rate $\nu_{ex}=\nu_{\infty}$ instead of $\nu_s$.

Thus, our theory reproduces the KF theory for $T \rightarrow T_c$ and the AG theory at $h_{\parallel} = 0$, as it should. It also formally reduces to the AG result in the fully polarized limit, $h_{\parallel}\rightarrow\infty$, when the PE and OE are neglected and the full exchange scattering rate is replaced by its effective value, $\nu_s \mapsto \nu_{\infty}$.

\section{Superconductivity enhancement by a parallel magnetic field}\label{sec: superconductivity enhancement arbitrary T}

In this section, we show that superconductivity enhancement appears not only as an increase in the critical temperature but also in other superconducting characteristics. We focus on the superfluid density and the spectral energy gap. We also show that the effect can be observed not only close to $T_c$ but also at arbitrary temperatures. 

To this end, we present analytical results for superconductivity enhancement by comparing the unpolarized and fully polarized limits, in which Dyson's equation \eqref{eq:symmetric part} has the same form as in the AG theory. Because Dyson's equation in these limits is formally equivalent to that in the AG theory, and because Eqs.\ \eqref{eq: self-consistency}--\eqref{eq: spectral gap} have the same form in both theories, we can directly apply the analytical results of the AG theory, which are summarized in Appendix~\ref{Appendix: AG theory}. 

In this section, we assume that both the PE and OE contributions are negligible, $\gamma_{\parallel}\rightarrow 0$. This condition is satisfied under the assumptions of Eqs.~\eqref{eq: weakness of PE} and \eqref{eq: weakness of OE}. This allows us to focus on the polarization mechanism responsible for superconductivity enhancement and to neglect pair-breaking effects, which tend to mask it.

To go beyond the AG-like regime and describe arbitrary magnetic fields, for which (i)~the pair-breaking effects are not negligible and (ii)~the full exchange scattering rate is an operator, we perform numerical calculations in Sec.~\ref{sec: numerics}.

Although the analytical treatment based on Eqs.\ \eqref{eq: u at zero field} and \eqref{eq: us fully polarized} does not describe the full magnetic-field range, it gives the maximum possible enhancement. This follows from comparing the case without polarization, $h_{\parallel}=0$ [Eq.\ \eqref{eq: u at zero field}], with the fully polarized case, $h_{\parallel} \rightarrow \infty$ [Eq.\ \eqref{eq: us fully polarized}]. The enhancement at arbitrary magnetic field can also be analyzed in the weak-exchange-scattering limit, $\nu_s \ll T_{c0}$; see Appendix~\ref{Appendix: small nus}.

The qualitative reason for superconductivity enhancement at $h_{\parallel} \rightarrow \infty$ relative to $h_{\parallel}=0$ is simple. In the AG theory \cite{Abrikosov1960}, superconductivity is suppressed as the exchange scattering rate increases. In the unpolarized limit, this rate contains two contributions: processes without spin flips, proportional to $\nu_z$, and processes with spin flips, proportional to $\nu_{\perp}$. In the fully polarized limit, all impurity spins are aligned along the $z$ axis, and spin-flip processes are frozen out. Therefore, only magnetic scattering without spin flips remains in this limit, with $ \nu_z (h_{\parallel} \rightarrow \infty) = \nu_{\infty}$. The effective exchange scattering rate is then reduced, $\nu_{\infty} < \nu_s$ [see Eq.\ \eqref{eq: nu ex}], and superconductivity is enhanced. According to Eq.\ \eqref{eq: nu ex}, the strength of the enhancement is determined by the factor $S/(S+1)$ and is largest for impurity spin $S=1/2$.

\subsection{Critical temperature and restoration of superconductivity}\label{sec: critical temperature and superconductivity restoration}
After discussing the qualitative picture, we turn to quantitative results that follow directly from the AG theory \cite{Abrikosov1960}. First, we connect the AG and KF theories by reproducing the critical-temperature result. We then determine the range of $\nu_s$ in which a magnetic field can restore superconductivity.

In the unpolarized and fully polarized limits, the Cooperon can be calculated explicitly from Eq.\ \eqref{eq:symmetric part}, or equivalently from Eq.\ \eqref{eq: Cooperon}:
\begin{equation}
   C_0(\varepsilon_{n}) = \begin{cases}
     1/(|\varepsilon_{n}| + \nu_s),  &h_{\parallel}  = 0, \\
     1/(|\varepsilon_{n}|+ \nu_\infty),  &h_{\parallel}  \rightarrow \infty.
    \end{cases}
\end{equation}
The critical temperature is then given by $T_c = T_{\mathrm{AG}}(\nu_{ex})$,
where $T_{\mathrm{AG}} (\nu_{ex})$ is the critical temperature in the AG theory with the effective scattering rate equal to $\nu_{ex}$. This critical temperature can be calculated from Eq.\ \eqref{eq: Tc KF theory}, giving the following transcendental equation:
\begin{equation}\label{eq: critical temperature AG}
      \ln (T_{c0}/T_{c})= \psi(1/2 + \nu_{ex}/2 \pi T_c) - \psi(1/2),
\end{equation} 
where $\psi(x)$ is the digamma function. 

In the limit $\nu_{ex} \gg 2 \pi T_c$, one can find the critical exchange scattering rate $\nu^{\mathrm{cr}}_{ex}$ at which superconductivity is completely destroyed:
\begin{equation}\label{eq: critical values SC extermination}
   \nu_{ex}^{\mathrm{cr}} = \begin{cases}
     \nu_{s}^{\mathrm{cr}} = \frac{\Delta_{0}}{2}, &h_{\parallel}  = 0, \\
    \nu_{\infty}^{\mathrm{cr}}=\frac{S+1}{S} \frac{\Delta_{0}}{2},  &h_{\parallel}  \rightarrow \infty,
    \end{cases}
\end{equation}
where $\Delta_{0} =\pi T_{c0}/ \gamma$ is the zero-temperature order parameter value at $\nu_{s} = 0$, and $\gamma \approx 1.78$ is Euler's constant.

One intriguing consequence of Eq.\ \eqref{eq: critical values SC extermination}, already noted in the KF theory, is that superconductivity can be restored by a magnetic field in the range $\nu^{\mathrm{cr}}_{s}<\nu_s <\nu^{\mathrm{cr}}_{\infty}$ \cite{kharitonovFeigelman}. In this range, superconductivity is absent at zero field, $h_{\parallel}=0$, but is restored at finite field and persists as $h_{\parallel} \rightarrow \infty$.

\subsection{Gapless-to-gapped transition}

In addition to reproducing the KF result for field-induced restoration of superconductivity, our theory predicts an effect that we call the ``gapless-to-gapped transition.'' This transition changes the spectral gap from $E_g=0$ below a critical field, $h_{\parallel}<h^{\mathrm{gap}}_{\parallel}$, to $E_g>0$ above it, $h_{\parallel}>h^{\mathrm{gap}}_{\parallel}$. To demonstrate this transition, we again compare the unpolarized and fully polarized limits and use the AG-theory result for the spectral gap $E_g$.
 
In the AG theory, the transition from the gapped to the gapless state occurs at the critical value ${\nu}^{\mathrm{gap}}_{ex} = \Delta_{0} \exp(- \pi/4)$. Because the exchange scattering rates differ in the unpolarized and fully polarized limits, the corresponding boundary values also differ:
\begin{equation}\label{eq: boundaries of the gapless-ti-gapped transition}
   \nu_{ex}^{\mathrm{gap}} = \begin{cases}
     \nu_s^{\mathrm{gap}}= \Delta_{0} \exp(- \pi/4), &h_{\parallel} = 0, \\
   \nu_\infty^{\mathrm{gap}} =\frac{S+1}{S} \Delta_{0} \exp(- \pi/4), &h_{\parallel}  \rightarrow \infty.
    \end{cases}
\end{equation}
 Equations~\eqref{eq: critical values SC extermination} and \eqref{eq: boundaries of the gapless-ti-gapped transition} show that both boundary values increase by the factor $(S+1)/S$ because of the polarization mechanism. This is a direct consequence of the magnetic-field-induced decrease in the exchange scattering rate, described by the same factor; see Eq.~\eqref{eq: nu ex}.

\begin{figure*}[t]
\includegraphics[width=1\textwidth]{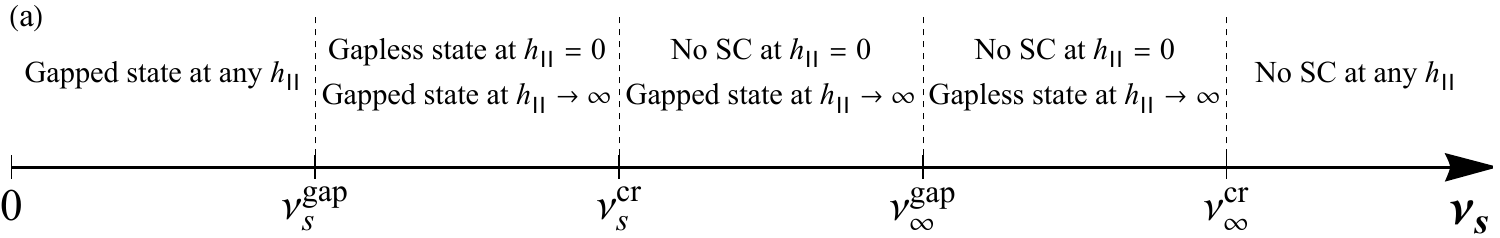} \\[5mm]
  \includegraphics[width=1\textwidth]{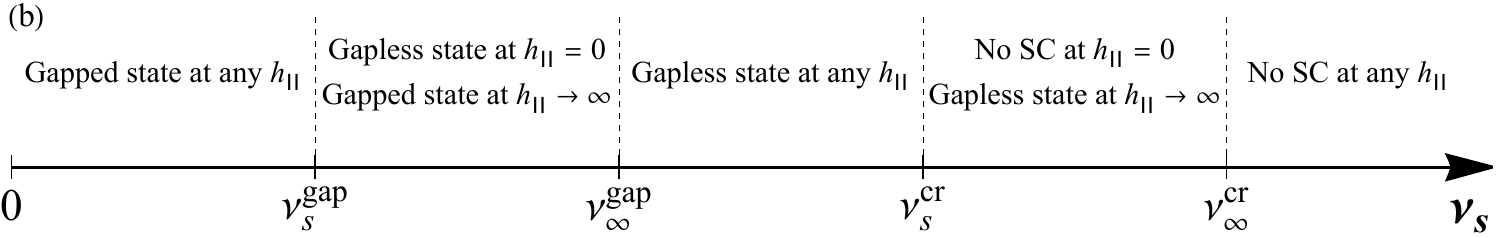}
\caption{Phase diagrams of the superconducting film for different exchange scattering rates $\nu_s$ and impurity spins: (a)~$S < S_0 \approx 10.3$ and (b)~$S > S_0$. The paramagnetic and orbital effects of the magnetic field are neglected. The gapless-to-gapped transition can only be observed in the range $\nu^{\mathrm{gap}}_s<\nu_s < \nu^{\mathrm{gap}}_{\infty}$, and magnetic-field-induced restoration of superconductivity (SC) is observed in the range $\nu^{\mathrm{cr}}_{s}<\nu_s < \nu^{\mathrm{cr}}_{\infty}$.}
 \label{fig: Phase diagram}
\end{figure*}

 Using Eqs.\ \eqref{eq: critical values SC extermination} and \eqref{eq: boundaries of the gapless-ti-gapped transition}, we can determine which field-induced transitions occur, neglecting the PE and OE, as functions of the zero-field exchange scattering rate $\nu_s$ and the impurity spin $S$. The resulting phase diagram is shown in Fig.\ \ref{fig: Phase diagram}.

For $S  <S_0 = 2/[\exp(\pi/4) -2] \approx 10.3 $ [Fig.~\ref{fig: Phase diagram}(a)], the boundary values satisfy $\nu_s^{\mathrm{cr}}<\nu^{\mathrm{gap}}_{\infty}$. The phase sequence is then as follows: (i)~for $\nu_s  < \nu_{s}^{\mathrm{gap}}$, both the unpolarized and fully polarized states are gapped; (ii)~for $\nu_{s}^{\mathrm{gap}} <\nu_s  < \nu_{s}^{\mathrm{cr}}$, the unpolarized state is gapless, whereas the fully polarized state is gapped; this is the gapless-to-gapped transition; (iii)~for $\nu_{s}^{\mathrm{cr}} <\nu_s  < \nu_{\infty}^{\mathrm{gap}}$, superconductivity is absent in the unpolarized state but restored as a gapped state in the fully polarized limit; (iv)~for $\nu_{\infty}^{\mathrm{gap}} <\nu_s  < \nu_{\infty}^{\mathrm{cr}}$, superconductivity is absent in the unpolarized state but gapless in the fully polarized state; and (v)~for $\nu_{\infty}^{\mathrm{cr}} <\nu_s$, superconductivity is absent in both limits.

For larger spins, $S  > S_0 $ [Fig.~\ref{fig: Phase diagram}(b)], the sequence changes: (i)~for $\nu_s  < \nu_{s}^{\mathrm{gap}}$, both the unpolarized and fully polarized states are gapped; (ii)~for $\nu_{s}^{\mathrm{gap}} <\nu_s  < \nu_{\infty}^{\mathrm{gap}}$, the unpolarized state is gapless, whereas the fully polarized state is gapped; (iii)~for $\nu_{\infty}^{\mathrm{gap}} <\nu_s  < \nu_{s}^{\mathrm{cr}}$, both states are gapless; (iv)~for $\nu_{s}^{\mathrm{cr}} <\nu_s  < \nu_{\infty}^{\mathrm{cr}}$, superconductivity is absent in the unpolarized state but gapless in the fully polarized state; and (v)~for $\nu_{\infty}^{\mathrm{cr}} <\nu_s$, superconductivity is absent in both limits.

Strictly speaking, the phase diagrams in Fig.~\ref{fig: Phase diagram} are valid only when the PE and OE are neglected. These effects are present in any real experiment and can modify the phase diagrams. In particular, they can make some phases unobservable, especially at large $S$, where the phase regions are narrower, scaling as $1/S$.

\subsection{Enhancement of the superfluid density}

Finally, we demonstrate the enhancement of the superfluid density $n_{sc}(\nu_{ex}, T)$. This requires solving Eq.\ \eqref{eq: nsc}, which relates the superfluid density to $u_s$, together with the self-consistency equation, Eq.\ \eqref{eq: self-consistency}. The corresponding AG-theory results are known for different magnetic-impurity strengths and temperature regimes; see Appendix~\ref{Appendix: AG theory}. For brevity, we focus on two limiting cases in which the results are analytical and simple: (i)~$\nu_{ex} \gg T_{\mathrm{AG}}(\nu_{ex})$ near the critical temperature $T_{\mathrm{AG}}(\nu_{ex})$, corresponding to strong suppression by magnetic impurities, and (ii)~$\nu_{ex} \ll T_{\mathrm{AG}}(0)= T_{c0}$ at $T=0$, corresponding to weak suppression by magnetic impurities.

Near $T_{\mathrm{AG}}(\nu_{ex})$, for $\nu_{ex} \gg T_{\mathrm{AG}}(\nu_{ex})$, one obtains \cite{Abrikosov1960}
\begin{equation}\label{eq: nsc near Tc}
    \frac{n_{sc} (\nu_{ex}, T)}{n} = \frac{8 \pi^2 T_{\mathrm{AG}}(\nu_{ex})[T_{\mathrm{AG}}(\nu_{ex}) - T]}{\nu_{0} \nu_{ex}}.
\end{equation}
At $T =0$ and for $\nu_{ex} \ll T_{c0}$, the superfluid density is \cite{Abrikosov1960}
\begin{equation}\label{eq:nsc at 0}
    \frac{n_{sc}(\nu_{ex},0)}{n} = \frac{\pi \Delta_0}{\nu_0} -  \left(\frac{4}{3} + \frac{\pi}{4} \right) \frac{\nu_{ex}}{\nu_0},
\end{equation}
where we have used the suppression of the order parameter, $\Delta = \Delta_0 - \pi\nu_{ex}/4$, obtained from the self-consistency equation.

Equations~\eqref{eq: nsc near Tc} and \eqref{eq:nsc at 0} show directly that the superfluid density is enhanced in the fully polarized limit relative to the unpolarized limit because $\nu_\infty < \nu_s$. They also show that this enhancement can occur at arbitrary temperature and for different magnetic-impurity strengths. The effect is especially pronounced near $T_c$ in the strong-magnetic-scattering regime. In this regime, the polarization mechanism can restore superconductivity, as discussed in Sec.\ \ref{sec: critical temperature and superconductivity restoration}, changing the superfluid density from zero to the finite value given by Eq.\ \eqref{eq: nsc near Tc}.

\section{Enhancement of the perpendicular upper critical field}\label{sec: enhancement of Hc2}

In Secs.\ \ref{sec: Diagrammatic technique} and \ref{sec: superconductivity enhancement arbitrary T}, we assumed that the applied magnetic field is parallel to the film surface, $\mathbf{h} = \mathbf{h}_{\parallel}$. Here we consider the more general case of a nonzero perpendicular component, $\mathbf{h}_{\perp} \neq 0$, and discuss the corresponding peculiarities of superconductivity enhancement.

The qualitative difference between parallel and perpendicular fields is due to the small film thickness $d$ compared with the in-plane sample size. Therefore, the orbital effect of the perpendicular field is usually much stronger than that of the parallel field, so that $\gamma_{\perp} \gg \gamma_{\parallel}$ and $H^{\perp}_{c2} \ll  H^{\parallel}_{c}$. A perpendicular field is therefore less favorable for observing polarization-induced enhancement of superconductivity. Nevertheless, signatures of the enhancement can still appear in the presence of a perpendicular field component.

To demonstrate this, we study the dependence of the perpendicular upper critical field on the parallel magnetic field, $h^{\perp}_{c2}(h_{\parallel})$, and show that it is nonmonotonic, with a maximum at a finite field. The parallel field, whose OE is weak, reduces the exchange scattering rate through the polarization mechanism and thereby enhances superconductivity. In particular, it enhances the perpendicular upper critical field $h^{\perp}_{c2}$.

To calculate $h^{\perp}_{c2}$, we generalize the Cooperon equation, Eq.\ \eqref{eq: Cooperon}, by including pair breaking due to the perpendicular field. Technically, we use a method similar to that introduced in Sec.~\ref{sec: dirty superconductor in the magnetic field} and calculate $C_0$ at a finite momentum $\mathbf{q}_s = e \mathbf{A}/c$. This gives the following equation for the Cooperon:
\begin{multline}\label{eq: Cooperon coordinate dep}
        \biggl[ |\varepsilon_n| + \hat{\nu}_{ex} +\gamma_{PE} + \frac{D}{2} \left(-i \frac{\partial}{\partial \mathbf{r}} - \frac{2 e \mathbf{A}}{c} \right)^2 \biggr] C_{0}(\varepsilon_n, \mathbf{r},\mathbf{r}')
        \\
        = \delta(\mathbf{r}-\mathbf{r}'),
\end{multline} 
where $\gamma_{PE} =  3h'^{2}/2\nu_{so}$ is the total PE depairing parameter and $h'^{2} = h'^2_{\parallel} + h'^{2}_{\perp}$. Equation~\eqref{eq: Cooperon coordinate dep} generalizes Eq.\ \eqref{eq: Cooperon} by explicitly including the real-space dependence of the Cooperon caused by the vector potential of a magnetic field with an arbitrary direction.

We then use the standard procedure \cite{deGennesBook, AGDBookEng, LevitovShytovBook} and expand the Cooperon in eigenfunctions $\varphi_{m}(\mathbf{r})$ of the diffusion operator $D \left(-i \partial/ \partial \mathbf{r} - 2 e \mathbf{A}/c \right)^2$; see Appendix~\ref{appendix: spatially dependent Cooperon} for details. The field $h^{\perp}_{c2}$ is determined by the lowest eigenvalue of this operator, so we keep only the zeroth mode. To solve the zeroth-mode equation, we assume a thin film, $d \ll l_{H}$, and average Eq.\ \eqref{eq: Cooperon coordinate dep} over the film thickness. The resulting in-plane problem has the form of a Schr\"odinger equation for a particle in a perpendicular magnetic field \cite{deGennesBook, AGDBookEng, LevitovShytovBook}. The Cooperon equation then has the same form as Eq.\ \eqref{eq: Cooperon}, but with the orbital pair-breaking parameter for a parallel field replaced by the total orbital pair-breaking parameter, $\gamma^{\parallel}_{OE} \mapsto \gamma_{OE}$, which includes the perpendicular-field contribution:
\begin{equation}\label{eq: orbitall effect}
\gamma_{OE}= \frac{2}{9}(p_{F}d)^2 \frac{h_{\parallel}^2}{\nu_0} + \frac{2 }{3} \left(p_{F} l \right) h_{\perp},
\end{equation}
The first term in Eq.\ \eqref{eq: orbitall effect} arises from averaging over the film thickness. This term was obtained in the KF theory and gives the orbital effect of the parallel field component, $\gamma^{\parallel}_{OE}$. The second term in Eq.\ \eqref{eq: orbitall effect} arises from the in-plane Schr\"odinger equation and gives the orbital effect of the perpendicular field component, $\gamma^{\perp}_{OE}$.

Finally, by solving Eqs.\ \eqref{eq: Cooperon} and \eqref{eq: orbitall effect} numerically, we obtain $h_{c2}^{\perp}(h_{\parallel})$. As mentioned above, this dependence shows enhancement due to the polarization effect caused by the parallel magnetic-field component. We demonstrate this explicitly in the next section.

\section{Numerical results: nonmonotonic behavior}\label{sec: numerics}

\begin{figure*}[t]
\includegraphics[width=0.66\columnwidth]{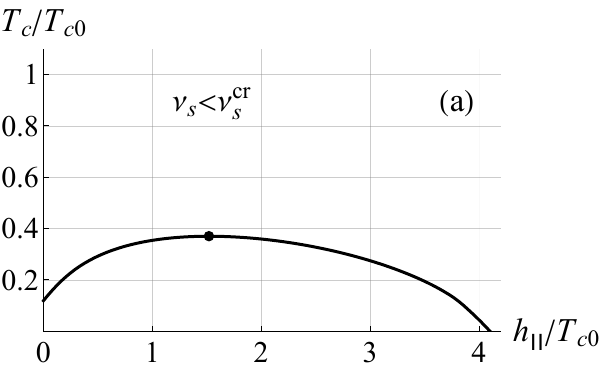}
 \hfill
  \includegraphics[width=0.66\columnwidth]{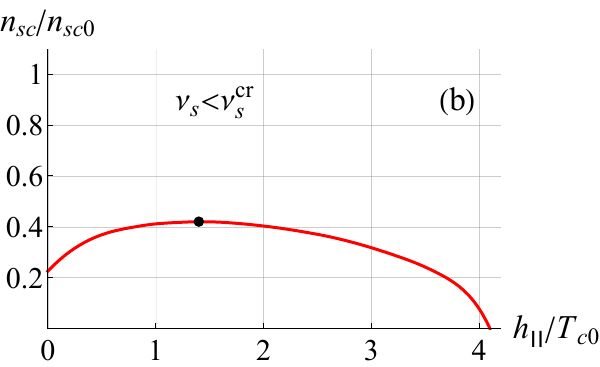} \hfill
  \includegraphics[width=0.66\columnwidth]{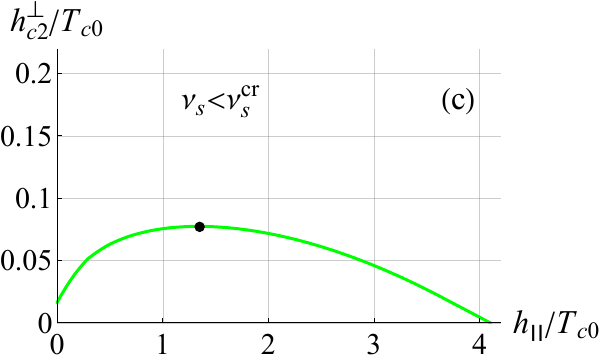} \\
  
 \includegraphics[width=0.66\columnwidth]{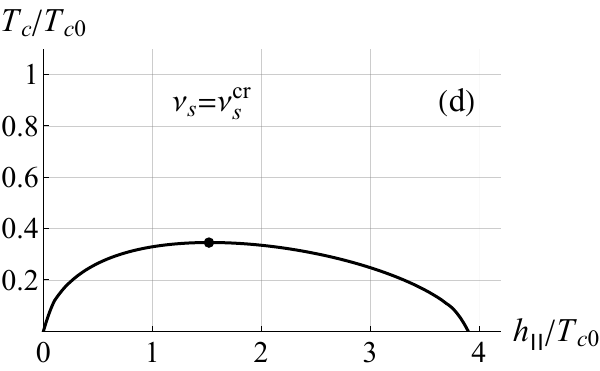}
 \hfill
  \includegraphics[width=0.66\columnwidth]{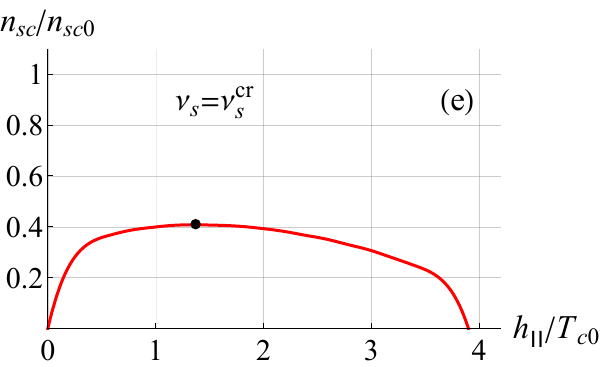} \hfill
  \includegraphics[width=0.66\columnwidth]{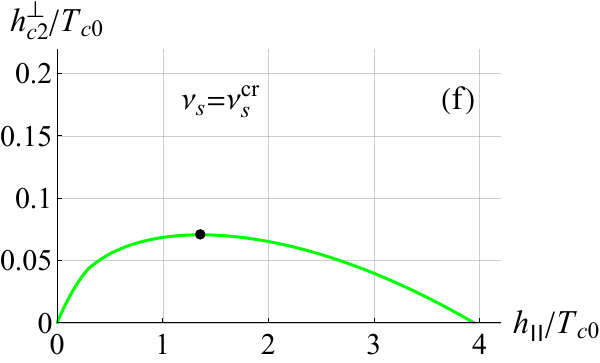} \\
  
 \includegraphics[width=0.66\columnwidth]{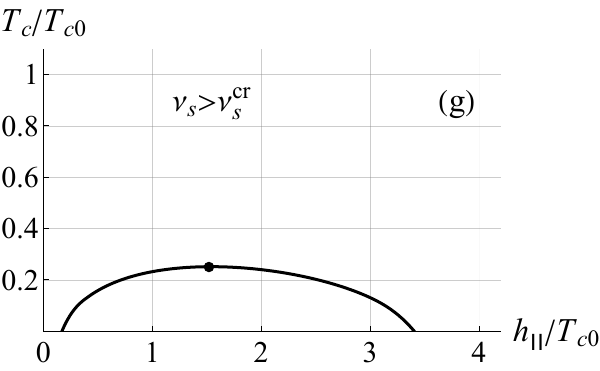}
 \hfill
  \includegraphics[width=0.66\columnwidth]{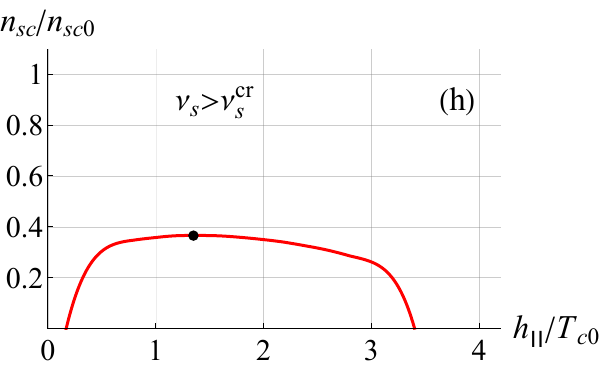} \hfill
  \includegraphics[width=0.66\columnwidth]{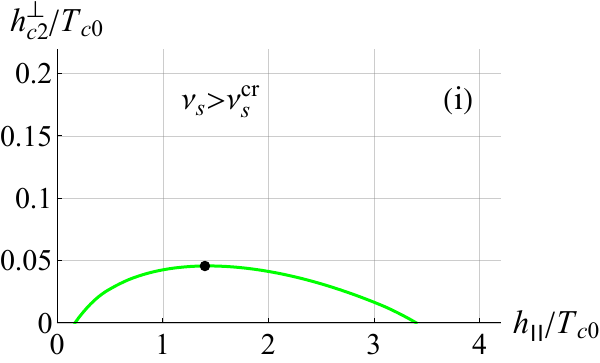} 
\caption{Nonmonotonic dependence of superconducting observables on the parallel magnetic field $h_{\parallel}$: (a), (d), and (g)~critical temperature $T_c$; (b), (e), and (h)~superfluid density $n_{sc}$ at $T = 0$; and (c), (f), and (i)~perpendicular upper critical field $h^{\perp}_{c2}$ at $T =0$. The superfluid density is normalized to $n_{sc0}$, its value at $h_{\parallel}=0$ and $\nu_s = 0$. The nonmonotonic curves have maxima at finite fields, marked by black dots, and arise from competition between the polarization and pair-breaking effects of the magnetic field. All panels are calculated for $S = 1/2$, $g_s=2$, $\nu_0 = 10^4 T_{c0}$, $\nu_{so} = 10^3 T_{c0}$, $\zeta_s = 5$, $p_{F}d = 30$, $p_{F}l = 5$, and different exchange scattering rates $\nu_{s}$. Panels~(a), (b), and~(c) correspond to $\nu_s = 0.85 T_{c0} <\nu^{\mathrm{cr}}_{s} \approx0.88 T_{c0}$, for which the state at $h_{\parallel}=0$ is superconducting. Panels~(d), (e), and~(f) correspond to $\nu_s = \nu^{\mathrm{cr}}_{s}$, for which the state at $h_{\parallel}=0$ is at the threshold of superconductivity, so that superconductivity is restored by any finite field. 
Finally, panels~(g), (h), and~(i) are obtained at $\nu_s = T_{c0} >\nu^{\mathrm{cr}}_{s}$, for which the state at $h_{\parallel}=0$ is normal and superconductivity is restored only at $h^{\mathrm{res}}_{\parallel} \approx 0.2 T_{c0}$.}
 \label{fig: Superconductivity enhancement}
\end{figure*}

We now complement the analytical results with numerical calculations. This allows us to describe the nonmonotonic behavior of superconducting observables caused by the competition between the polarization effect and magnetic-field pair breaking. In this section, we include finite PE and OE, so that the conditions in Eqs.\ \eqref{eq: weakness of PE} and \eqref{eq: weakness of OE} need not be satisfied. As examples, we study the dependences of the critical temperature $T_{c}$, the superfluid density $n_{sc}$, and the perpendicular upper critical field $h^{\perp}_{c2}$ on the parallel field $h_{\parallel}$. The results for different exchange scattering rates $\nu_s$ are shown in Fig.~\ref{fig: Superconductivity enhancement}. The curves for $T_c$ are obtained from the KF theory, whereas those for $n_{sc}$ and $h^{\perp}_{c2}$ follow from our extended model.

All superconducting characteristics shown in Fig.~\ref{fig: Superconductivity enhancement} are nonmonotonic functions of the parallel magnetic field. Depending on $\nu_s$, the zero-field state can be superconducting, for $\nu_s < \nu^{\mathrm{cr}}_{s}$, or normal, for $\nu_s > \nu^{\mathrm{cr}}_{s}$. If the zero-field state is superconducting, then all considered observables are nonzero at $h_{\parallel}=0$, as shown in Figs.~\ref{fig: Superconductivity enhancement}(a)--\ref{fig: Superconductivity enhancement}(c). The polarization effect then enhances already existing superconductivity. If the zero-field state is normal, then the observables vanish at $h_{\parallel}=0$, but the polarization effect can restore superconductivity at a finite field $h^{\mathrm{res}}_{\parallel}$. This restoration field increases with $\nu_s$. It starts from zero at $\nu_s = \nu^{\mathrm{cr}}_s$, as shown in Figs.~\ref{fig: Superconductivity enhancement}(d)--\ref{fig: Superconductivity enhancement}(f), and becomes finite for $\nu_s > \nu^{\mathrm{cr}}_s$, as shown in Figs.~\ref{fig: Superconductivity enhancement}(g)--\ref{fig: Superconductivity enhancement}(i).

In all cases in Fig.~\ref{fig: Superconductivity enhancement}, the polarization effect increases the superconducting observables, either from finite zero-field values or from zero. As the magnetic field increases, however, magnetic-field pair breaking also grows. The resulting competition between polarization, which enhances superconductivity, and pair breaking, which suppresses it, produces maxima at finite fields. At larger fields, pair breaking dominates and superconductivity is suppressed. Eventually, all superconducting observables in Fig.~\ref{fig: Superconductivity enhancement} vanish at a critical field, signaling the destruction of superconductivity by depairing.

\section{Comparison with experiment}\label{sec: Comparison with the experimental results}

\begin{figure*}[t]
\includegraphics[width=0.67\columnwidth]{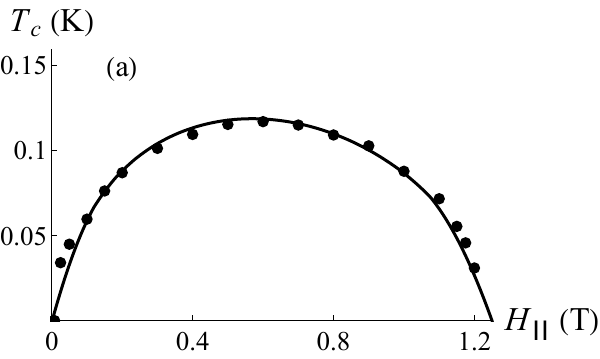}
 \hfill
  \includegraphics[width=0.67\columnwidth]{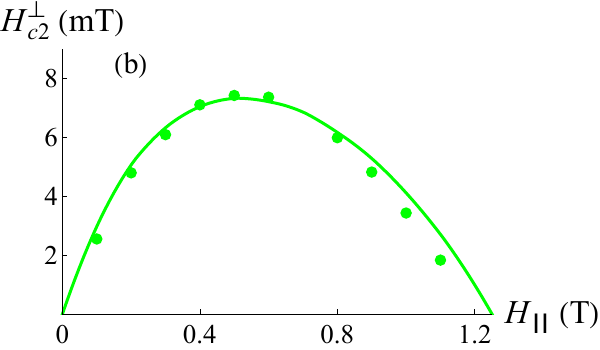} \hfill
  \includegraphics[width=0.67\columnwidth]{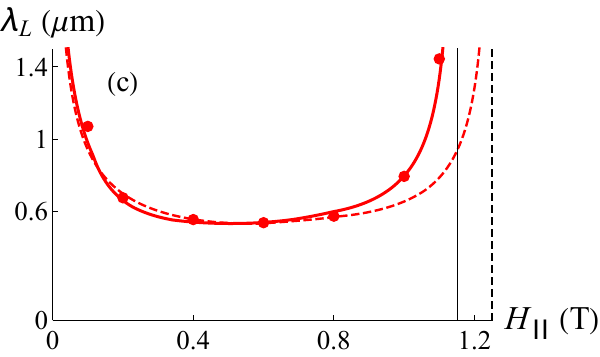} \\
  
\caption{Comparison between the experimental data of Ref.\ \cite{Llanos2026} (dots) and the corresponding theoretical curves (solid and dashed lines) as functions of $H_{\parallel}$ for (a)~the critical temperature $T_c$, (b)~the perpendicular upper critical field $H^{\perp}_{c2}$, and (c)~the London penetration depth $\lambda_L$. For all theoretical curves, we use $S = 5/2$, $g_s=2$, $\nu_0 = 1032 T_{c0}$, $\nu_{so} = 348 T_{c0}$, $\nu_s = 0.89 T_{c0}$, $\zeta_s = 1.25$, $p_{F}d = 20$, and $T_{c0} = 717$\,mK. These values are taken from Table~S4 of Ref.\ \cite{Llanos2026}. Panel~(a) follows from the KF theory \cite{kharitonovFeigelman, Llanos2026}, whereas the fits in panels~(b) and~(c) are obtained from our extended theory. The fits in panels~(b) and~(c) require additional parameters. We use $p_{F} l = 14$ for panel~(b). For the dashed theoretical curve in panel~(c), we use $\lambda_{L0}= 0.33$\,$\mu$m for the London penetration depth in the absence of a magnetic field and magnetic impurities. This dashed curve does not include the depairing effect of the perpendicular magnetic field required by the experimental protocol for measuring $\lambda_L$. This omission leads to a deviation from the experimental data near $H^{\parallel}_c \approx1.25$\,T, shown by the dashed black asymptote. Including the additional depairing $\gamma^{\perp}_{OE}=0.015 T_{c0}$ gives the solid curve, which provides a better fit, especially near the additionally suppressed value $H^{\parallel}_{c} \approx 1.15$\,T shown by the solid black asymptote. For this fit, we use $\lambda_{L0}= 0.29$\,$\mu$m.}
 \label{fig: Comparison with experiment}
\end{figure*}

In this section, we compare the results of our extended theoretical model with the experimental data of Ref.\ \cite{Llanos2026}. That work studied thin superconducting LaSb$_{2}$ films doped with magnetic Ce atoms and subjected to an external magnetic field. By varying the concentration of magnetic impurities, the authors changed the exchange scattering rate $\nu_s$ and explored the polarization effect in different regimes. In particular, they reached both the superconductivity-enhancement regime, $\nu_s < \nu^\mathrm{cr}_s$, and the superconductivity-restoration regime, $\nu_s > \nu^\mathrm{cr}_{s}$; see Fig.~2(c) and Fig.~S11 of Ref.\ \cite{Llanos2026}. 

In Ref.\ \cite{Llanos2026}, the authors measured the dependences $T_c(H_{\parallel})$ for different values of $\nu_s$ and found good agreement with the KF theory. In addition, at $\nu_{s} = 0.89 T_{c0}$, they directly measured the perpendicular upper critical field $H^{\perp}_{c2}$ and indirectly extracted the London penetration depth $\lambda_L$ by fitting the data to a collective vortex-creep model. These measurements were performed for different values of the parallel field component $H_{\parallel}$. The experimental results and theoretical fits are shown in Fig.~\ref{fig: Comparison with experiment}.

Figure~\ref{fig: Comparison with experiment}(a) shows the KF-theory result, which agrees well with the experimental data \cite{kharitonovFeigelman, Llanos2026}. The results of our extended theory are shown in Figs.~\ref{fig: Comparison with experiment}(b) and~\ref{fig: Comparison with experiment}(c). The curve in Fig.~\ref{fig: Comparison with experiment}(b) also agrees well with the experiment. In fitting Fig.~\ref{fig: Comparison with experiment}(b), we use the same parameter set as in Fig.~\ref{fig: Comparison with experiment}(a),
with one additional parameter, $p_{F}l$, which enters the orbital pair-breaking parameter in Eq.\ \eqref{eq: orbitall effect}.

Figure~\ref{fig: Comparison with experiment}(c) shows the London penetration depth $\lambda_L$, which we calculate from the superfluid density $n_{sc}$ as \cite{deGennesBook, TinkhamBook, AbrikosovBookEng}
\begin{equation}
    \lambda_L = \sqrt{mc^2/4\pi n_{sc}e^2}.
\end{equation}
Using the same parameter set as in Fig.~\ref{fig: Comparison with experiment}(a), we obtain the dashed curve in Fig.~\ref{fig: Comparison with experiment}(c). This curve, however, deviates from the experiment near the critical field  $H^{\parallel}_c$, where $T_c$ and $H^{\perp}_{c2}$ vanish in Figs.~\ref{fig: Comparison with experiment}(a) and~\ref{fig: Comparison with experiment}(b), respectively. We attribute this deviation to the experimental protocol used to measure $\lambda_L$. In the experiment, a perpendicular magnetic field was needed to create vortices, so that $\lambda_L$ could be inferred from their collective creep. This perpendicular field also produces an additional orbital pair-breaking parameter $\gamma^{\perp}_{OE}$, which is not included in the dashed-curve fit in Fig.~\ref{fig: Comparison with experiment}(c). This parameter suppresses $H^{\parallel}_{c}$ and modifies the theoretical curve near the critical field. We therefore treat $\gamma^{\perp}_{OE}$ as an additional fitting parameter and obtain the solid curve in Fig.~\ref{fig: Comparison with experiment}(c). This curve agrees well with the experiment over the full range of $H_{\parallel}$.

Thus, our results indicate that the nonmonotonic behavior of $H^{\perp}_{c2}(H_{\parallel})$ and $\lambda_L (H_{\parallel})$ observed in Ref.\ \cite{Llanos2026} has the same origin as the nonmonotonic behavior of $T_{c}(H_{\parallel})$. In all three cases, the behavior results from competition between the polarization effect induced by the parallel field component and pair breaking produced by both the parallel and perpendicular field components.

\section{Discussion}\label{sec:discussion}
\subsection{Disorder in the experiment}

The description of the polarization effect within the KF theory, and likewise within our extended model, assumes the presence of potential, spin-orbit, and magnetic impurities. In the experiment of Ref.\ \cite{Llanos2026}, the magnetic impurities can be clearly attributed to Ce doping. The sources of potential and spin-orbit scattering are less obvious, especially because the films were obtained by epitaxial growth.

The most plausible explanation is that nonmagnetic scattering is caused by surface scattering. In particular, the fit of $H^{\perp}_{c2}(H_{\parallel})$ in Fig.~\ref{fig: Comparison with experiment}(b) agrees best with the experiment for $l/d \approx 0.7$. This fitted value is consistent at the order-of-magnitude level with an estimate based on the potential-scattering rate obtained from the KF fit in Ref.\ \cite{Llanos2026}. Indeed, assuming the effective mass to be equal to the bare electron mass, $m=m_e$, we would then obtain $l/d\approx1.2$. This indicates that the mean free path is of the order of the film thickness, supporting the surface-scattering interpretation. Experimental estimates of the mean free path from Hall measurements \cite{Llanos2026} give $l/d \approx 3$, which is also in rough agreement with this interpretation. However, the value of $l/d$ used in the fit in Fig.~\ref{fig: Comparison with experiment}(b) gives much better agreement with the experiment.

Alternatively, both potential and spin-orbit scattering could be attributed to scattering by the same Ce atoms. However, the very low concentration of these atoms can hardly explain the relatively high potential- and spin-orbit-scattering rates. Spin-orbit scattering could also arise from coupling between potential scattering and the intrinsic spin-orbit interaction associated with heavy atoms in the LaSb$_2$ film, through the Elliott--Yafet mechanism \cite{Boross2013}.

\subsection{Role of dimensionality and Fermi-surface structure in LaSb$_2$}

The KF theory and our extended model are formulated for a single isotropic three-dimensional band with a spherical Fermi surface in the diffusive regime (including diffusive motion across the film thickness). These assumptions are only marginally controlled for the LaSb$_2$ films. Their electronic structure is likely multiband and anisotropic \cite{Llanos2026}, while the fit in Fig.~\ref{fig: Comparison with experiment}(b) gives $l/d \approx 0.7$ instead of the asymptotic condition $l/d\ll1$.

Fermi-surface anisotropy, multiband effects, and electronic dimensionality may therefore quantitatively modify the effective diffusion coefficients and scattering rates entering the theory. In most cases, we do not expect these changes to alter the qualitative structure of the results. An important exception is orbital pair breaking by a parallel magnetic field, which is particularly sensitive to the film dimensionality: its contribution is proportional to $d^2$ and vanishes in the strict $2$D limit. For the experimental film \cite{Llanos2026}, however, our fit gives $l/d\approx0.7$, placing the system on the diffusive side of the diffusive-to-ballistic crossover rather than in the ballistic regime $l/d\gg 1 $. Although the asymptotic condition $l/d \ll1$ is not parametrically satisfied, the fitted value is consistent with using the finite-thickness diffusive model as an effective description.

Most importantly, the qualitative polarization mechanism does not rely on the dimensionality or detailed shape of the Fermi surface. It originates from the polarization of the magnetic-impurity spins and the associated freeze-out of spin-flip processes. Fermi-surface geometry affects the quantitative values of the effective scattering rates but not the mechanism itself. Consequently, the single-band isotropic diffusive model captures the qualitative physics and also provides a good quantitative description demonstrated in Fig.~\ref{fig: Comparison with experiment}.
\color{black}

 \subsection{Outlook: experimental suggestions}

We suggest several directions for future experimental studies of the polarization effect. The results could then be compared with our theory.

First, we suggest extracting the London penetration depth $\lambda_L$, and therefore the superfluid density $n_{sc}$, from kinetic-inductance measurements rather than from collective vortex-creep measurements in a finite perpendicular field. This would avoid the unwanted pair-breaking effect of the auxiliary perpendicular field, which is not intrinsically required to study the London penetration depth.

Second, the predicted gapless-to-gapped transition could be tested by tunneling spectroscopy. In this case, we also expect a nonmonotonic dependence of the spectral gap $E_g$ on the parallel magnetic field, with a shape similar to that of $T_c(H_{\parallel})$.

\section{Conclusions}\label{sec:conclusions}
We have investigated a thin dirty superconducting film containing potential, spin-orbit, and magnetic impurities in an external magnetic field. For this system, KF predicted a mechanism of superconductivity enhancement caused by polarization of the impurity spins, which reduces the effective exchange scattering rate $\nu_{ex}$. However, the original work \cite{kharitonovFeigelman} described only the dependence of the critical temperature $T_c$ on a magnetic field parallel to the film surface, $h=h_{\parallel}$. Motivated by the recent experiment of Ref.\ \cite{Llanos2026}, we extended the KF theory to arbitrary temperatures and magnetic-field orientations using Gor'kov's diagrammatic technique for dirty superconductors.

For a magnetic field parallel to the film surface, $h=h_{\parallel}$, we generalized the description to arbitrary temperatures below the critical value, $T<T_c$. We derived Dyson's equation for the Green's function including the reduced exchange scattering rate and presented a general scheme for calculating superconducting observables.

To analyze the superconductivity enhancement analytically, we considered the unpolarized ($h\ll T_{c0}$) and fully polarized ($h\gg T_{c0}$) limits while neglecting magnetic-field pair breaking. In both limits, our theory maps onto the AG theory with an effective exchange scattering rate $\nu_{ex}$. As in the KF theory, the fully polarized rate, $\nu_{ex}=\nu_{\infty}$, is smaller than the unpolarized rate, $\nu_{ex}=\nu_s$, so that $\nu_{\infty}<\nu_s$. This reduction is caused by the freeze-out of spin-flip scattering on magnetic impurities and explains the enhancement relative to the unpolarized case. We showed that the enhancement can increase the spectral gap $E_g$, so that an applied magnetic field can induce a gapless-to-gapped transition. We also demonstrated explicitly that the magnetic field increases the superfluid density $n_{sc}$. 

We also generalized the KF theory to an arbitrary field orientation. For simplicity, we restricted this analysis to the vicinity of $T_c$. We modified the Cooperon equation to include the orbital effect of the perpendicular field component $h_{\perp}$. The theory then predicts enhancement of the perpendicular upper critical field $h^{\perp}_{c2}$ due to polarization induced by the parallel field component.

To describe the competition between polarization and magnetic-field pair breaking, we performed numerical calculations within the extended theory. We obtained the dependences $n_{sc}(h_{\parallel})$ and $h_{c2}^{\perp}(h_{\parallel})$ for several values of $\nu_s$. The resulting curves are nonmonotonic and exhibit maxima at finite field, reflecting this competition. Depending on $\nu_s$, we found two regimes of field-induced enhancement: enhancement of an existing superconducting state for $\nu_s<\nu_s^{\mathrm{cr}}$, and restoration of superconductivity for $\nu_s>\nu_s^{\mathrm{cr}}$.

Finally, we compared the predictions of our theoretical model for the dependences of the perpendicular upper critical field $h^{\perp}_{c2}$ and the London penetration depth $\lambda_L$ on the parallel magnetic field with the experimental data of Ref.\ \cite{Llanos2026}. Our theory agrees well with the experimental results. This supports the interpretation that the observed nonmonotonic behavior results from competition between the polarization effect caused by the parallel field component and pair breaking produced by both the parallel and perpendicular field components. 

\acknowledgments
We thank M.~V.\ Feigel'man for drawing our attention to this problem and for useful discussions of the results.
We are also grateful to L.~R.\ Tagirov and I.~V.\ Tokatly for useful comments and discussion.
We thank R.\ Dorrian and J.\ Falson for discussion of their experimental data.


\appendix

\section{Irrelevance of disorder-induced interference}\label{appensix: interferential terms}

Our assumption of independent disorder differs from that of the KF theory, where potential scattering is coupled to spin-orbit and magnetic scattering. In our notation, the KF assumption would mean that (i)~$H_{eU}$ in Eq.\ \eqref{eq: HeU} is combined with $H_{eSO}$ in Eq.\ \eqref{eq: HeSO}, with the impurity sum running over the same index ``$b$'', and (ii)~$H_{eS}$ in Eq.\ \eqref{eq: HeS} contains an additional potential part at the same impurity sites labeled by the index ``$c$''.

However, interference between different scattering mechanisms does not appear in the Cooperon equation and therefore does not affect the critical temperature $T_c$ \cite{kharitonovFeigelman}. There are two reasons for this: (i)~the interference term arising from spin-orbit and potential scattering vanishes after angular integration, and (ii)~the interference term arising from magnetic and potential scattering reduces to a spin-dependent shift of the chemical potential, which effectively disappears after momentum integration (the so-called $\xi_p$ integration) in each spin subband.  

We emphasize that in many other cases the interference between various types of disorder also does not affect physical observables \cite{Abrikosov1960, Gorkov1964, Tsuneto1964}. In some specific cases, however, such interference can be important and can lead to new effects \cite{Sauls2003, Shen2014, Huang2018, Burmistrov2025}.

\section{Derivation of Dyson's equation}
\label{Appendix: Dyson's equation}
In this Appendix, we give details of the derivation of Dyson's equation. This requires calculating the self-energy diagrams in Fig.~\ref{fig: Diagramms} in the self-consistent Born approximation, using the exact Green's function $\hat{G}^{\mathbf{q}_s}_{\pm}$ given by Eq.\ \eqref{eq:Green function zero perp field}. The resulting equations for the renormalized frequencies $\tilde{\varepsilon}_{n}$, magnetic field $\tilde{h}$, and superconducting gaps $\tilde{\Delta}_{\pm}$ take the form
\begin{widetext}
\begin{multline}\label{eq: self consistence for epsilon}
    \tilde{\varepsilon}_n \pm i \tilde{h}_{\parallel} =  \varepsilon_{n} \pm ih'_{\parallel}  + \frac{\nu_0 +\nu_{so}/3 + \nu_z}{2}\frac{\tilde{\varepsilon}_n\pm i \tilde{h}_{\parallel}}{\sqrt{  (\tilde{\varepsilon}_n \pm i \tilde{h}_{\parallel} )^2+\tilde{\Delta}_{\pm}^2}} + \frac{\nu_{so}}{3} \frac{\tilde{\varepsilon}_n \mp i \tilde{h}_{\parallel}}{\sqrt{  (\tilde{\varepsilon}_n \mp i \tilde{h}_{\parallel} )^2+ \tilde{\Delta}_{\mp}^2}}  \\ 
    +\frac{\nu_{0} v_{F}^2 \langle q_s^2\rangle_d }{4}\frac{( \tilde{\varepsilon}_n \pm i \tilde{h}_{\parallel}) \tilde{\Delta}_{\pm}^2 }{\bigl[  (\tilde{\varepsilon}_n \pm i \tilde{h}_{\parallel} )^2+ \tilde{\Delta}_{\pm}^2\bigl]^{5/2}}+\frac{\nu_{s} \langle S_z\rangle  T}{S (S+1)} \sum\limits_{\omega_{m}} \frac{\omega_{s} \mp i \omega_{m}}{\omega_{s}^2 + \omega_{m}^2} \frac{ \tilde{\varepsilon}_n \mp i \tilde{h}_{\parallel}}{\sqrt{(\tilde{\varepsilon}_n \mp i \tilde{h}_{\parallel} )^2+ \tilde{\Delta}_{\mp}^2}}\Biggl|_{\varepsilon_{n} - \omega_{m}},  
\end{multline}
and 
\begin{multline}\label{eq: self consistence for Delta}
    \tilde{\Delta}_{\pm} = \Delta + \frac{\nu_0 +\nu_{so}/3 - \nu_z}{2}\frac{\tilde{\Delta}_{\pm}}{\sqrt{(\tilde{\varepsilon}_{n} \pm i \tilde{h}_{\parallel} )^2 + \tilde{\Delta}_{\pm}^2 }} + \frac{ \nu_{so}}{3} \frac{\tilde{\Delta}_{\mp}}{\sqrt{(\tilde{\varepsilon}_{n} \mp i \tilde{h}_{\parallel} )^2 + \tilde{\Delta}_{\mp}^2 }} \\
   -
    \frac{\nu_{0}v_{F}^2 \langle q_s^2\rangle_d  }{12}\frac{[ 2( \tilde{\varepsilon}_{n} \pm i \tilde{h}_{\parallel})^2-\tilde{\Delta}_{\pm}^2 ] \tilde{\Delta}_{\pm} }{\bigl[(\tilde{\varepsilon}_{n}\pm i \tilde{h}_{\parallel})^2+ \tilde{\Delta}_{\pm}^2 \bigl]^{5/2}}-\frac{\nu_{s} \langle S_z \rangle T}{ S (S+1)} \sum\limits_{\omega_{m}} \frac{\omega_{s} \mp i \omega_{m}}{\omega_{s}^2 + \omega_{m}^2} \frac{\tilde{\Delta}_{\mp}}{\sqrt{(\tilde{\varepsilon}_{n} \mp i \tilde{h}_{\parallel} )^2+ \tilde{\Delta}_{\mp}^2 }}\Biggl|_{\varepsilon_{n} - \omega_{m}}. 
\end{multline}
Here, $|_{\varepsilon_n - \omega_m}$ denotes that the arguments of the functions under the sum are shifted, i.e., one should take $\tilde{\varepsilon}_n (\varepsilon_n- \omega_m)$, $\tilde{h}_{\parallel} (\varepsilon_n- \omega_m)$, and $\tilde{\Delta}_{\mp} (\varepsilon_n- \omega_m)$. The quantity $\langle q^2_s \rangle_d$ is the average of $\mathbf{q}_s^2$ over the film thickness.

Different terms in Eqs.\ \eqref{eq: self consistence for epsilon} and \eqref{eq: self consistence for Delta} originate from different self-energy diagrams, have different structures, and describe different effects. First, terms that mix variables with different spin indices ($\pm$ signs) correspond to spin-flip processes caused by spin-orbit and magnetic impurities. Second, pair-breaking effects appear in these equations when $h_{\parallel} \neq 0$ and $\nu_s \neq 0$.

The pair-breaking effects can be captured by a single equation following from Eqs.\ \eqref{eq: self consistence for epsilon} and \eqref{eq: self consistence for Delta}. This equation is conveniently written in terms of the functions $u_{\pm} = (\tilde{\varepsilon}_{n} \pm i \tilde{h}_{\parallel})/\tilde{\Delta}_{\pm}$ and takes the form
\begin{equation}\label{eq:general equation}
    \frac{\varepsilon_{n} \pm i h'_{\parallel}}{\Delta} = u_{\pm}\left(1 - \frac{\nu_{z} + \gamma^{\parallel}_{OE}}{\Delta} \frac{1}{\sqrt{1 + u^2_{\pm}}} \right)  + \frac{\nu_{so}}{3 \Delta} \frac{u_{\pm} - u_{\mp}}{\sqrt{1 + u_{\mp}^2}} - \frac{\nu_{s} \langle S_z\rangle T}{  S (S+1)\Delta}   \sum \limits_{\omega_{m}} \frac{ \omega_{s} \mp i \omega_{m}}{\omega_s^2 + \omega_{m}^2} \frac{u_{\pm }+\bar{u}_{\mp}}{\sqrt{1 + \bar{u}_{\mp}^2}} ,
\end{equation}
\end{widetext}
where $\gamma^{\parallel}_{OE} = 2v_{F}^2 \langle q_s^2 \rangle_d /3\nu_0$ is the exact orbital pair-breaking parameter, which leads to the expression in Eq.\ \eqref{eq: pair-breaking parameters}, and $\bar{u}_{\pm}= u_{\pm}(\varepsilon_n - \omega_m)$. Pair breaking leads to a deviation of $u_{\pm}$ from the value $\varepsilon_n/\Delta$. 

Finally, we decompose $u_{\pm}$ into symmetric and antisymmetric parts, $u_{\pm}=u_s\pm u_a$. The antisymmetric component $u_a\neq0$ describes the paramagnetic effect and is induced by a finite Zeeman field, $h_{\parallel}\neq0$. In the strong spin-orbit-scattering regime, however, the paramagnetic effect is suppressed, so that $u_a$ is parametrically smaller than $u_s$ (by a factor $h'_{\parallel}/ \nu_{so}$). One may therefore proceed perturbatively: first determine $u_a$ and then calculate the corresponding correction to $u_s$. To the leading order, the equation for $u_a$ is obtained by retaining only the Zeeman term, proportional to $h'$, in the left-hand side of Eq.\ \eqref{eq:general equation} and the spin-orbit term, proportional to $\nu_{\mathrm{so}}$, in its right-hand side, which yields Eq.\ \eqref{eq:asymmetric part}.

Equation~\eqref{eq:asymmetric part} illustrates that the spin-orbit scattering does not have the pair-breaking effect. On the contrary, it  suppresses the paramagnetic effect and thereby protects superconductivity. The qualitative reason is that the spin-orbit scattering preserves the time-reversal symmetry. At the same time, by mixing spin-up and spin-down states, the spin-orbit scattering reduces the effective influence of the Zeeman field.

Finally, by considering the symmetric part of Eq.\ \eqref{eq:general equation} in the leading order with respect to the spin-orbit scattering, we recover Eq.\ \eqref{eq:symmetric part}.

\section{Results of the AG theory}
\label{Appendix: AG theory}

In this Appendix, we recall the main results of the AG theory used in our analysis of the unpolarized and fully polarized limits. The AG theory was originally derived for a superconductor containing magnetic impurities \cite{Abrikosov1960}. Dyson's equation can then be written in the form of Eq.\ \eqref{eq: u at zero field}. The self-consistency equation, the expression for the superfluid density $n_{sc}$, and the density of states $N$ are given by Eqs.\ \eqref{eq: self-consistency}--\eqref{eq: spectral gap}.

\subsection{Self-consistency equation}

The self-consistency equation can be analyzed at $T \rightarrow T_c$ and $T =0$. While $T_c$ itself is determined by Eq.\ \eqref{eq: critical temperature AG}, at $T \rightarrow T_{c}$ we can expand Eq.\ \eqref{eq: self-consistency} up to the second order with respect to small $\Delta$:
\begin{multline}\label{eq: self cons with Delta2}
    \ln\frac{T_{c0}}{T}= \psi\left(\frac{1}{2} + \rho_{s}\right) - \psi\left(\frac{1}{2}\right) \\ +\left(\frac{\Delta}{4 \pi T}\right)^2 \left[\psi^{(2)}\left(\frac{1}{2} + \rho_s \right) + \frac{\rho_{s}}{3} \psi^{(3)}\left(\frac{1}{2} + \rho_s \right)\right],
\end{multline}
where $\rho_s = \nu_s/2\pi T$, whereas $\psi(x)$ is the digamma function and $\psi^{(n)}(x)$ is its $n$th derivative.  
From Eq.\ \eqref{eq: self cons with Delta2}, we obtain the following temperature dependence of the order parameter
\begin{multline}\label{eq: Delta near Tc}
    \Delta^2(T) = (4\pi T_{c})^2 \left( 1 - \frac{T}{T_{c}} \right)  \left[1 - \rho_{c}\psi^{(1)}\left(\frac{1}{2} + \rho_{c}\right)\right]\\ \times
    \left[\psi^{(2)}\left(\frac{1}{2} + \rho_c \right) + \frac{\rho_{c}}{3} \psi^{(3)}\left(\frac{1}{2} + \rho_c \right)\right]^{-1},
\end{multline}
where $\rho_c = \nu_s/2 \pi T_{c}$. It can be further simplified if $\nu_{s} \gg T_c$:
\begin{equation}\label{eq: Delta Tc2}
    \Delta^2(T) = 4 \pi^2 T_c (T_c - T).
\end{equation}

At zero temperature, $T=0$, replacing the sum in Eq.\ \eqref{eq: self-consistency} by an integral gives the following transcendental equation
for the order parameter
\begin{equation}\label{eq: Delta at T=0}
    \ln \frac{\Delta_0}{\Delta} = \begin{cases}
        \pi  \eta/4 , & \eta \leq 1, \\
        \arccosh \eta + \frac{1}{2} \left(\eta \arcsin\frac{1}{\eta}  -\sqrt{1- \frac{1}{\eta^2}}\right), & \eta > 1,
    \end{cases}
\end{equation}
where $\eta = \nu_s/\Delta$. The solution of Eq.\ \eqref{eq: Delta at T=0} determines the dependence of $\Delta$ on $\nu_s$ at $T = 0$. In the limiting case $\nu_{s} \ll \Delta_0$, one finds
\begin{equation}\label{eq: Delta at small nu_s}
\Delta  = \Delta_{0} - \pi \nu_{s}/4.
\end{equation}

\subsection{Superfluid density}
Next, we discuss the AG-theory results for the superfluid density $n_{sc}$ in the dirty limit $\nu_0 \gg \Delta_0$. To this end, we analyze Eqs.\ \eqref{eq: nsc} and \eqref{eq: u at zero field} at different temperatures.

At $ T \rightarrow T_{c}$, we can expand the equations in powers of $u_{s} \gg 1$, which leads to 
\begin{equation}\label{eq: superfluid density at Tc1}
    n_{sc}/n = [\Delta^2(T)/ \pi \nu_0 T_c] \psi^{(1)}(1/2 + \rho_{c}).
\end{equation}
It can be further simplified if $\nu_s \gg T_c$:
\begin{equation}\label{eq: superfluid density at Tc2}
    n_{sc}/n = \Delta^2(T)/ \pi \nu_0 T_c \rho_c .
\end{equation}
Equations \eqref{eq: Delta Tc2} and \eqref{eq: superfluid density at Tc2} result in Eq.\ \eqref{eq: nsc near Tc}.

In the opposite limit, $T=0$, we replace the sum in Eq.\ \eqref{eq: nsc} by an integral and obtain
\begin{widetext}
\begin{equation}\label{eq: superfluid density zero temperature}
    \frac{n_{sc}}{n} = \frac{\Delta}{\nu_0 } \times \begin{cases}
        \pi - 4 \eta/3 , & \eta \leq 1,\\ 
        2\arcsin {\frac{1}{\eta}} - 2\eta \left[ \frac{2}{3} -  \sqrt{1-\frac{1}{\eta ^2}} +\frac{1}{3} \left(1-\frac{1}{\eta ^2}\right)^{3/2}\right], & \eta > 1.
    \end{cases}
\end{equation}
\end{widetext}
Taking the limit $\nu_s \ll \Delta_0$ in Eq.\ \eqref{eq: superfluid density zero temperature} and using Eq.\ \eqref{eq: Delta at small nu_s}, we reproduce Eq.\ \eqref{eq:nsc at 0}.
 
\subsection{Density of states}
One of the central results of the AG theory is the prediction of gapless superconductivity. This state still has superconducting correlations but, unlike a gapped state, has a finite density of states $N(E)$ at any (real) energy $E$; in other words, it has no energy gap. To study the density of states, one must analytically continue Eq.\ \eqref{eq: u at zero field} to real energies. This gives 
\begin{equation}\label{eq: real freq}
E/\Delta =u - \eta u/{\sqrt{1 - u^2}} ,
\end{equation}
where $u(E)$ is the analytical continuation of $u_s(\varepsilon_n)$. The density of states in Eq.\ \eqref{eq: spectral gap} can then be rewritten as
\begin{equation}\label{eq: density of states AG}
    N(E) = (N_F/\eta)\Im u.
\end{equation}

Equation~\eqref{eq: density of states AG} implies that the density of states $N(E)$ vanishes when $u$ is a real solution of Eq.\ \eqref{eq: real freq} at a given energy $E$. This real solution must also satisfy $\sgn u(E) = \sgn E$. For $\eta < 1$, such a real solution exists for any $E< E_g$, where $E_g$ is the spectral gap. It is obtained by maximizing the right-hand side of Eq.\ \eqref{eq: real freq}, which gives
\begin{equation}
    E_g = \Delta \bigl( 1 - \eta^{2/3} \bigr)^{3/2}.
\end{equation}
For $\eta \geq 1$, no such real solution of Eq.\ \eqref{eq: real freq} exists, implying $E_g = 0$. Thus, $\eta = 1$ corresponds to the transition from the gapped to the gapless state. To obtain the boundary value $\nu_s^{\mathrm{gap}}$ for this transition, one can use the first line of Eq.\ \eqref{eq: Delta at T=0} at $\eta = 1$. The result is given by the first line of Eq.\ \eqref{eq: boundaries of the gapless-ti-gapped transition}.

\section{Limit of weak exchange scattering}\label{Appendix: small nus}

As a supplement to Sec.\ \ref{sec: superconductivity enhancement arbitrary T}, we now consider another limiting case in which superconductivity enhancement can be analyzed analytically. This is the weak magnetic-scattering limit, $\nu_s \ll T_{c0}$. In this limit, the exchange-scattering terms in Eq.\ \eqref{eq:symmetric part} are small and can be treated perturbatively. This allows us to analyze superconductivity enhancement at an arbitrary parallel magnetic field. For simplicity, as in Sec.\ \ref{sec: superconductivity enhancement arbitrary T}, we neglect the PE and OE, setting $\gamma_{\parallel} = 0$, and focus on the polarization mechanism.

We solve the problem perturbatively in $\nu_s$. To this end, we write $u_s$ as the series
\begin{equation}
    u_s = u^{(0)}_s +  u^{(1)}_{s} + \dots 
\end{equation}
In the zeroth order of the perturbation theory, we neglect the terms proportional to $\nu_s$ in  Eq.\ \eqref{eq:symmetric part} and obtain
\begin{equation}
    u^{(0)}_{s} = \varepsilon_{n}/\Delta.
\end{equation}
 In the first order, we substitute $u^{(0)}_s$ into the term proportional to $\nu_s$ and obtain

\begin{multline}\label{eq: us1 general}
u^{(1)}_{s} = \frac{\nu_{z}}{\Delta} \frac{\varepsilon_{n}}{\sqrt{ \varepsilon_{n}^2+\Delta^2 }} \\+ \frac{\nu_{s} \langle S_z \rangle T}{S (S+1) \Delta} \sum_{\omega_m} \frac{ \omega_{s}}{\omega_{s}^2+ \omega_m^2}\frac{2 \varepsilon_{n} - \omega_{m} }{\sqrt{(\varepsilon_{n} - \omega_{m})^2+\Delta^2 }}. 
\end{multline}
Below, we consider the zero-temperature limit $T = 0$, where the sum can be replaced by an integral.

\subsection{Effective exchange scattering rate}
\begin{figure}[t!]
\includegraphics[width=\columnwidth]{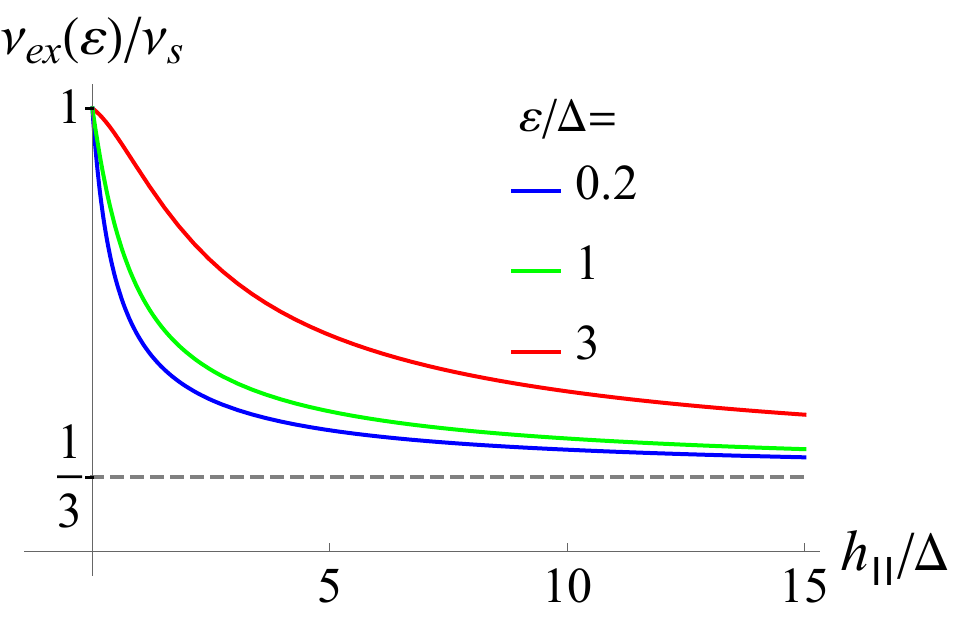}
\caption{Effective exchange scattering rate $\nu_{ex}(\varepsilon)$ [see Eq.\ \eqref{eq: u1}] at different Matsubara frequencies $\varepsilon$ as a function of the parallel magnetic field $h_{\parallel}$ at $T = 0$, $S = 1/2$, and $\nu_{s} \ll T_{c0}$. As the magnetic field increases, the effective exchange scattering rate decreases because of the polarization effect. In the fully polarized limit $h_{\parallel} \rightarrow \infty$, it approaches its minimum value $\nu_s S/(S+1)$, which equals $\nu_s/3$ for $S=1/2$. The decay rate decreases as $\varepsilon$ increases.}\label{fig: scattering}
\end{figure}

At $T=0$, Eq.\ \eqref{eq: us1 general} can be written in a form similar to the AG result:
\begin{equation}
    u^{(1)}_{s}  = \frac{\nu_{ex}( \varepsilon)}{\Delta}\frac{\varepsilon}{\sqrt{\varepsilon^2+ \Delta^2}},
\end{equation}
by introducing the effective exchange scattering rate $\nu_{ex} (\varepsilon)$. As a function of $x = \varepsilon/\Delta$, it has the form
\begin{multline}\label{eq: u1}
    \nu_{ex}(x)=  \nu_z \\ +\frac{\nu_{s} \langle S_z \rangle}{S (S+1)} \frac{\sqrt{1+x^2}}{x}\int_{-\infty}^{\infty} \frac{dt}{2\pi} \frac{ x_{s}}{x_{s}^2+ (t-x)^2}\frac{x +t }{\sqrt{1 + t^2}},
\end{multline}
where $x_s = \omega_s/\Delta$.

We emphasize that the effective exchange scattering rate $\nu_{ex}$ depends on the Matsubara frequency $\varepsilon$, which becomes a continuous variable at $T=0$. In the AG theory, by contrast, the corresponding rate $\nu_{ex} (\varepsilon) = \nu_s$ is frequency independent. Therefore, the quantity $\nu_{ex}(\varepsilon)$ introduced above is not an actual AG exchange scattering rate. It is nevertheless a useful measure of the strength of exchange scattering.

As shown in Fig.~\ref{fig: scattering}, the effective exchange scattering rate decreases monotonically with increasing $h_{\parallel}$. This decrease is caused by the polarization of magnetic impurities. The decay rate becomes smaller as $\varepsilon$ increases.

\subsection{Enhancement of the superfluid density}

\begin{figure}[t!]
\includegraphics[width=\columnwidth]{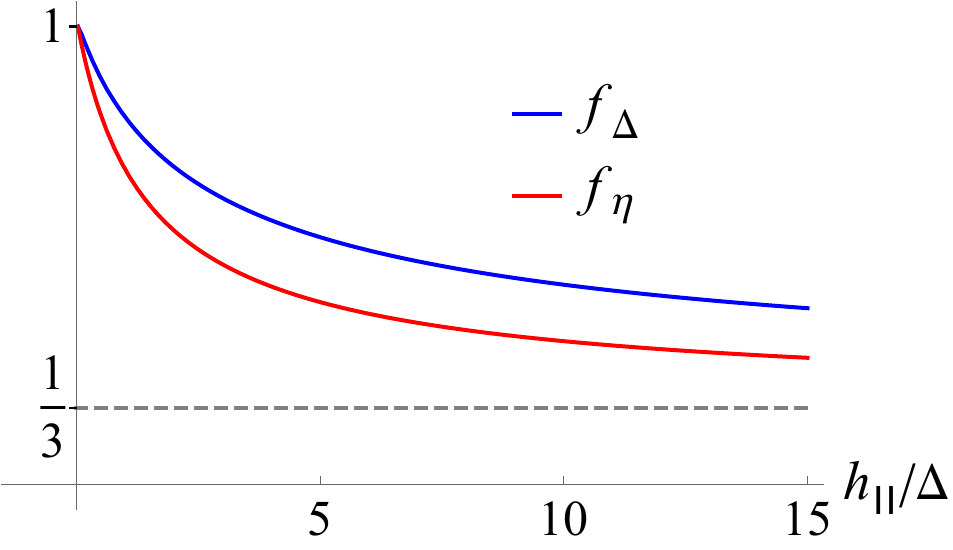}
\caption{Dependence of the renormalization factors $f_{\Delta}$ and $f_{\eta}$ [defined by Eqs.\ \eqref{eq: fDelta at small nus} and \eqref{eq: fnsc at small nus}, respectively] on the parallel magnetic field $h_{\parallel}$ at $T = 0$, $S = 1/2$, and $\nu_s \ll T_{c0}$. As the magnetic field increases, both factors decrease monotonically, indicating the enhancement of $\Delta$ and $n_{sc}$. In the fully polarized limit $h_{\parallel} \rightarrow \infty$, both factors tend to $S/(S+1)$, which equals $1/3$ for $S=1/2$.}
\label{fig: fdelta}
\end{figure}

Using Eq.\ \eqref{eq: u1}, we can now calculate the superfluid density and demonstrate its enhancement. We first solve the self-consistency equation, Eq.\ \eqref{eq: self-consistency}, perturbatively in the weak-exchange-scattering limit:
\begin{gather}\label{eq: Delta appendix}
    \Delta = \Delta_0-(\pi \nu_{s}/ 4)  f_{\Delta}, \\
    f_{\Delta} =\frac{4}{\pi}\int_{0}^{\infty} d x \frac{u^{(1)}_{s} x}{(1+x^2)^{3/2}} . 
    \label{eq: fDelta at small nus}
\end{gather}
Comparing with the AG result in Eq.\ \eqref{eq: Delta at small nu_s}, we can interpret $f_{\Delta}$ as the renormalization of the pair-breaking parameter for $\Delta$.

Finally, using Eqs.\ \eqref{eq: nsc} and \eqref{eq: Delta appendix}, we calculate the superfluid density:
\begin{gather}\label{eq: nsc at small nus}
    \frac{n_{sc}}{n} = \frac{\pi \Delta_{0}}{\nu_0} - \frac{\nu_s}{\nu_0} \left(\frac{\pi^2}{4}f_{\Delta} + \frac{4}{3}f_{\eta} \right), \\
    f_{\eta} =  3\int_{0}^{\infty} dx \frac{ u^{(1)}_{s} x}{(1+x^2)^2}.
    \label{eq: fnsc at small nus}
\end{gather}
Comparing with the AG result given in the first line of Eq.\ \eqref{eq: superfluid density zero temperature}, we can interpret $f_{\eta}$ as the renormalization of $\eta$ entering the expression for $n_{sc}$. The total suppression of the superfluid density includes both the $f_{\Delta}$ and $f_{\eta}$ contributions, as shown in Eq.\ \eqref{eq: nsc at small nus}. 

Both renormalization factors are shown in Fig.~\ref{fig: fdelta}. They behave similarly and decrease monotonically as the magnetic field increases. This is a consequence of the polarization effect. The reduction of both factors enhances the order parameter and the superfluid density, in accordance with Eqs.\ \eqref{eq: Delta appendix} and \eqref{eq: nsc at small nus}.

\section{Spatially dependent Cooperon}\label{appendix: spatially dependent Cooperon}
Because of the orbital effect of the magnetic field associated with the vector potential $\mathbf{A}$, the Cooperon equation for $C_0$ becomes spatially dependent; see Eq.\ \eqref{eq: Cooperon coordinate dep}. The self-consistency equation becomes spatially dependent as well and is given by
\begin{equation}\label{self-constency coordinate dep}
    \Delta(\mathbf{r}) = \lambda N_F \pi T \sum_{\varepsilon_{n}} \int C_0(\varepsilon_{n},\mathbf{r}, \mathbf{r}') \Delta(\mathbf{r}') d\mathbf{r}'.
\end{equation}

Therefore, to solve this self-consistency equation and obtain the upper critical field $h^{\perp}_{c2}$, one must find the spatial dependence of the Cooperon. We use the standard technique \cite{deGennesBook, AGDBookEng, LevitovShytovBook} and expand the Cooperon and the order parameter in eigenfunctions $\varphi_{m}(\mathbf{r})$ of the diffusion operator $D \left(-i \partial/ \partial \mathbf{r} - 2 e \mathbf{A}/c \right)^2$:
\begin{gather}
C_0 (\varepsilon_n,\mathbf{r},\mathbf{r}') = \sum\limits_{m=0}^{\infty}C^{(m)}_{0}(\varepsilon_{n}) \varphi_{m}^{*}(\mathbf{r}) \varphi_{m}(\mathbf{r}'), \\\Delta(\mathbf{r}) = \sum\limits_{m=0}^{\infty} \Delta^{(m)} \varphi_{m}(\mathbf{r}).
\end{gather}
The resulting coefficients $C_0^{(m)}$ and $\Delta^{(m)}$ satisfy spatially independent equations:
\begin{gather}\label{eq: Cooperon with perpendicular magnetic field}
        \left(|\varepsilon_n| + \hat{\nu}_{ex}+\gamma_{PE} + \gamma^{(m)} \right) C^{(m)}_{0}(\varepsilon_n) = 1,\\
    \Delta_{m} = \lambda N_F \pi T \sum_{\varepsilon_n} C^{(m)}_0( \varepsilon_{n}) \Delta_{m},\label{eq: self-consistency Cooperon with perpendicullar field}
\end{gather}
where $\gamma^{(m)}$ is the $m$th eigenvalue of the diffusion operator. 

As explained in the main text, $h^{\perp}_{c2}$ is determined by the lowest eigenvalue $\gamma^{(0)}$, which takes the form given by Eq.\ \eqref{eq: orbitall effect}.

\end{document}